\documentclass[showpacs, amsmath, amsfonts, amssymb, twocolumn,superscriptaddress,pra]{revtex4}
\usepackage{graphicx}
\usepackage{color}
\usepackage[normalem]{ulem}

\usepackage{ulem}  

\begin{document}

\newcommand{\vect}[1]{{\mathrm {\mathbf #1}}} 
\newcommand{\real}[1]{{\mathrm Re}\, #1} 
\newcommand{\imag}[1]{{\mathrm Im}\, #1} 

\title{Thick-medium model of transverse pattern formation in optically excited cold two-level atoms with a feedback mirror}

\author{W.~J.~Firth\footnote{To whom correspondence should be addressed.}\email[]{w.j.firth@strath.ac.uk}}
\author{I. Kre\v si\' c\footnote{Now at: Institute of Physics, Bijeni\v{c}ka cesta 46, 10000 Zagreb, Croatia}}

\affiliation{SUPA and Department of Physics, University of
Strathclyde, 107 Rottenrow East, Glasgow G4 0NG, UK}

\author{G. Labeyrie}

\author{A. Camara}

\affiliation{Universit\'{e} C\^{o}te d'Azur, CNRS, Institut de Physique de Nice, 06560
Valbonne, France}

\author{T. Ackemann}
\affiliation{SUPA and Department of Physics, University of
Strathclyde, 107 Rottenrow East, Glasgow G4 0NG, UK}

\date{\today}

\begin{abstract}
We study a pattern forming instability in a laser driven optically thick cloud of cold two-level atoms with a planar feedback mirror. A theoretical model is developed, enabling a full analysis of transverse patterns in a medium with saturable nonlinearity, taking into account diffraction within the medium, and both the transmission and reflection gratings. Focus of the analysis is on combined treatment of nonlinear propagation in a diffractively- and optically-thick medium and the boundary condition given by feedback. We demonstrate explicitly how diffraction within the medium breaks the degeneracy of Talbot modes inherent in thin slice models. Existence of envelope curves bounding all possible pattern formation thresholds is predicted. The importance of envelope curves and their interaction with threshold curves is illustrated by experimental observation of a sudden transition between length scales as mirror displacement is varied.

\end{abstract}

\maketitle

\section{Introduction}

Self-organization of light and atomic degrees of freedom in laser driven systems of cold atoms with optical feedback has in recent years received considerable attention \cite{ritsch13}. In addition to the longitudinal axis (e.g. of an optical cavity), spatial ordering can also occur in the plane transverse to the driving laser beam.

Transverse optical self-organization has been studied in a wide range of non-linear media during the last 30 years \cite{cross93, arecchi99}. A particularly simple and fruitful setup is the single feedback mirror (SFM) configuration, where a non-linear medium experiences double-pass excitation by a single single pump beam with mirror feedback. Spatial coupling of tranversely separate regions inside the medium is provided by diffraction \cite{firth90a, ackemann01}. Recently, we have used this setup to observe long-range hexagonal ordering in a thermal cold atomic gas, breaking the continuous spatial symmetries of the initial system \cite{labeyrie14,Camara2015}. This matches interest in a related scheme for patterns in cold atom systems interacting with two independent counterpropagating input fields \cite{Muradyan2005,Greenberg2011a, Schmittberger2016, Greenberg2011b,labeyrie16}.

Employing cold atoms as optical media offers a high degree of tunability such that the mechanism of the optical non-linearity can be selected by e.g.\ the duration of the pump pulse. For long pulses ($>10\, \mu$s), with blue detuning, optomechanical \cite{Bjorkholm1978,ashkin82} density modulations were shown to be dominant in optimum conditions \cite{labeyrie14}, whereas for shorter pulses ($<2\,\mu$s), pattern formation was found to be consistent with the standard two-level electronic nonlinearity \cite{Camara2015}. The results of Ref. \cite{Camara2015} constitute the first observation of pattern formation in a system with a saturable electronic two-level nonlinearity.

As was highlighted in our earlier work, the full analysis of both qualitative and quantitative features of the transverse patterns in cold atoms demands a departure from the ``thin-medium" approximation, in which diffraction within the medium is assumed negligible in comparison with the free-space diffraction between the medium and the mirror. One goal of the present paper is to derive a new, ``thick-medium", model of the two-level instability with the inclusion of diffraction within the nonlinear medium and to investigate how its predictions compare to experimental results.

A major advance from previous models of the SFM configuration is the inclusion of diffraction within the optical medium. The requisite theory is related to that used to analyze pattern formation in a mirrorless thick-medium (slab) with two counterpropagating (CP) input fields. Such CP systems have been analyzed for Kerr media by Firth et al \cite{firth90b} and Geddes et al \cite{Geddes1994}, and by Muradyan et al \cite{Muradyan2005}, as part of a study of optomechanical
effects in cold atoms.

Our model also includes the simultaneous presence of transmission gratings (purely transverse gratings resulting from the interference of the pump with copropagating sidebands) and reflection gratings (wavelength scaled gratings which result from the interference of counterpropagating beams) in the presence of feedback mirror, whereas earlier treatments only utilized pure transmission gratings \cite{firth90a,dalessandro92,ackemann95b}. Two-beam coupling via pure reflection gratings was included in the analysis of photorefractive experiments \cite{Honda1996}.

A system somewhat analogous to the present one was studied in Ref.~\cite{kozyreff06}, where dispersion in the time domain plays the role of diffraction in the spatial domain. The analogy is limited, however, because the interacting beams are co- and not counter-propagating, which leads to analytical differences. More important, reflection gratings, crucial in the cold-atom SFM and CP systems, are necessarily absent from the system analyzed in Ref.~\cite{kozyreff06}.

A key advance in the present paper is that we also include a full treatment of absorption (and its saturation), not included in the above-mentioned works. This is necessary to treat the region of small pump detuning, where absorptive effects were seen to limit pattern formation in recent experiments~\cite{Camara2015}. There is no known analytic solution to the thick-medium threshold equations in the presence of absorption, but we have developed an efficient and instructive graphical approach to the numerical evaluation of threshold curves. A side-benefit of our approach is our demonstration that, as the feedback mirror distance is varied, all the corresponding threshold curves are bounded by one or more envelope curves. These are as easily calculated as any single threshold curve, and are thus a very effective means of establishing the existence and extent of instability domains. Furthermore, we show that the zero-diffraction intercepts of these envelopes correspond exactly to thin-medium-model thresholds. This correspondence, the existence of envelope curves in SFM models, and our graphical ``gain-circle" approach to numerical evaluation of thresholds are likely to be applicable to SFM and related problems in a wide variety of nonlinear optical media.

\begin{figure}
\resizebox{85mm}{!}{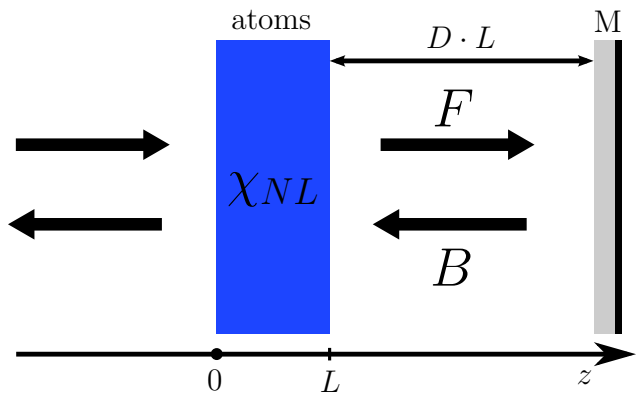}
\caption{ \label{fig:setup}
(Color online) Schematic of the SFM configuration. A linearly-polarized beam is sent into an atomic cloud modeled as a thick slab of length $L$ (blue online) with a non-linear susceptibility $\chi_{NL}$. The transmitted beam is retro-reflected by a mirror (M) with an adjustable displacement $D L$ beyond the end of the medium. The forward ($F$) and backward ($B$) propagating beams interfere inside the cloud. Experimental parameters: cloud of $^{87}$Rb atoms at $T=200\,\mu$K driven at a detuning of $\delta>0$ to the $F=2\to F'=3$ transition of the D$_2$-line, optical density (base $e$) in line center OD=210, effective sample size (FWHM of cloud) $L=8.5$~mm \cite{Camara2015}.
}
\end{figure}


\section{System and Model}
\label{sec_model}

Figure \ref{fig:setup} shows a schematic of our setup. A medium of length $L$ is illuminated by a laser beam leading to a forward field $F$. The transmitted light is retro-reflected by a plane mirror leading to a backward field $B$. We are scaling the longitudinal coordinate by the medium length $L$. Hence the normalized feedback distance $D$ measured from the exit face of the medium to the mirror is $D L$ in units of distance. (The mirror distance $d$ used in \cite{Camara2015} is measured from medium centre, $d=(D+1/2)L$.)

Similar to Muradyan et al \cite{Muradyan2005}, which we will refer to as MM, we consider the counter-propagating fields $F$ and $B$ to be coupled by a nonlinear susceptibility

\begin{equation}
\chi_{NL}= - \frac{6\pi}{k_0^3} n_a \frac{2\delta/\Gamma -i}{1+4\delta^2/\Gamma^2}\frac{1}{1+I/I_{s\delta}}
\label{chiNL}
\end{equation}
Here $n_a$ is the atomic density (considered constant here). $I$ is the intensity, which will be a
standing wave: $I/I_{s\delta} = |Fe^{ikz} + B e^{-ikz}|^2$. We can conveniently rewrite (\ref{chiNL}) as
\begin{equation}
\chi_{NL}= \chi_l \frac{1}{1+I/I_{s\delta}}
\label{chisat}
\end{equation}
where $\chi_l$ is the linear susceptibility (and is complex, though absorption is neglected in the MM, making the system Kerr-like).

As in MM, we use a time-independent susceptibility approach to the two-level nonlinearity. This precludes consideration of growth rates or oscillatory instabilities \cite{LeBerre1991}, but leads to reasonably tractable and transparent models which allow the parameter dependences of pattern thresholds to be investigated. We include absorption, so as to allow for arbitrary atom-field detunings. We include reflection-grating to all orders (MM include such effects, but only at lowest order). This analysis will be applied to the calculation of thresholds for transverse instability in the full thick-medium two-level model in Sections \ref{sec:transpert} and subsequent. Various limits and approximations of the full model will be discussed, so as to connect with earlier work. These include the Kerr limit, used for the thick-medium calculations presented in Fig. 3B of  \cite{labeyrie14}. In \cite{Camara2015} preliminary two-level results were presented for two cases: quasi-Kerr (i.e.\ large detuning, neglecting absorption, but not saturation of the refractive nonlinearity) for the pattern size vs mirror displacement; and absorptive thin-slice for the threshold vs atomic detuning.

The next step is to expand the nonlinear factor in a  Fourier series:

\begin{equation}
\frac{1}{1+I/I_{s\delta}}= \sigma_0 +\sigma_+ e^{2ikz} +\sigma_-e^{-2ikz} + h.o.t.
\label{Igrat}
\end{equation}

The higher-order  terms  do not lead to any phase matched couplings, and so can reasonably be neglected whatever the intensity. The coefficients $\sigma_{\pm}$ evidently describe a $2k$ longitudinal modulation of the susceptibility, i.e.\ a reflection (Bragg) grating, which will  scatter the forward field into the backward one and vice versa.

The field equations (M3) of \cite{Muradyan2005} can then be written as

\begin{equation}
\label{modeleqs}   \left \{
\begin{array}{l}
\frac{\partial F}{\partial z} - \frac{ i}{2k}\nabla^2_\perp F=i \frac{k}{2}\chi_l(\sigma_0 F +\sigma_+ B) ,\\
\\ \frac{\partial B}{\partial z} + \frac{ i}{2k}\nabla^2_\perp B = -i \frac{k}{2}\chi_l(\sigma_- F +\sigma_0 B) \\
\end{array} \right.
\end{equation}

To calculate $\sigma_{0,\pm}$, we write the exact expansion of the saturation term (\ref{Igrat}) as
\begin{equation}
\frac{1}{1+I/I_{s\delta}}= \frac{1}{1+p+q} (1+r(e_{+} + e_{+}^{*}))^{-1}
\label{gratexp}
\end{equation}
where $|F(z)|^2 = p(z)$,  $|B(z)|^2 = q(z)$ and  $e_{+} = e^{2ikz} e^{i(\theta_F-\theta_B)}$, with
$\theta_{F,B} = \text{arg}(F,B)$.

We have introduced a coupling parameter $r=h(pq)^\frac{1}{2}/(1+p+q)$, where the ``grating parameter" $h$ \cite{firth90b} allows consistent consideration of the cases of no reflection grating ($h=0$), and of a full grating ($h=1$). In the  former case $\sigma_{\pm}=0$, which would correspond to the standing-wave modulation of the susceptibility being washed out by drift or diffusion. Partial wash-out could be accommodated by intermediate values of $h$, but would need some associated physical justification. The MM model includes the full grating, so corresponds to $h=1$.

The series expansion of $(1+r(e_{+} + e_{+}^{*}))^{-1}$ is always convergent, because $r< 1/2$. Even terms contribute to $\sigma_0$, odd terms to $\sigma_{\pm}$. Using the binomial theorem, we find
\begin{equation}
\label{sigmas}   \left \{
\begin{array}{l}
(1+p+q)\sigma_0 = 1 +2r^2 +6r^4 + 20r^6 + ...\\
\\ (1+p+q)\sigma_+ = - e^{i(\theta_F-\theta_B)}(r + 3r^3 + 10r^5 + ...) \\
\end{array} \right.
\end{equation}
with $\sigma_- =  \sigma_+^{*}$.

The series in  (\ref{sigmas}) can be summed,  leading to a set of field evolution equations:

\begin{equation}
\label{exacteqs}   \left \{
\begin{array}{l}
\frac{\partial F}{\partial z} - \frac{ i}{2k}\nabla^2_\perp F =i \frac{k}{2}\chi_lF\left(\frac{1-\left(1-4r^2\right)^{-1/2}}{2hp}+\frac{\left(1-4r^2\right)^{-1/2}}{1+p+q}\right) ,\\
\\ \frac{\partial B}{\partial z} + \frac{ i}{2k}\nabla^2_\perp B = -i \frac{k}{2}\chi_l B\left(\frac{1-\left(1-4r^2\right)^{-1/2}}{2hq}+\frac{\left(1-4r^2\right)^{-1/2}}{1+p+q}\right) \\
%
%
%
\end{array} \right.
\end{equation}

Several papers, going back to the 1970s, have obtained analytic solutions (in the plane-wave limit) to (\ref{exacteqs}). For our purposes, the papers of van Wonderen et al  \cite{vanW1989,vanWonderen91} (who were addressing optical bistability in a Fabry-Perot cavity) are most directly relevant, and underpin the analytic zero-order (no diffraction) solution obtained in the next section.

For finite $h$, there is explicit nonreciprocity, since the susceptibilities for $F$ and $B$ are different, because of the susceptibility grating.  Quantitatively, the nonreciprocity is entirely due to the denominator, respectively $2hp$ and $2hq$, of the first term in the brackets on the right of (\ref{exacteqs}), the other terms all being symmetric in $p$ and $q$.  In the limit of no grating, $h,r \to 0$, both brackets reduce to the expected saturation denominator $(1+s)$, where the total intensity $s=p+q$. Even with a susceptibility grating present, the amplitudes $F$ and $B$ are slowly varying in $z$, allowing the propagation in the medium to be approximated by comparatively few longitudinal spatial steps.

In all the cases discussed above, we can write the two propagation equations in the form

\begin{equation}
\label{transeqs}   \left \{
\begin{array}{l}
\frac{\partial F}{\partial z} - \frac{ iL}{2k}\nabla^2_\perp F =-\frac{\alpha_lL}{2}(1+i\Delta) A(p,q)F ,\\
\\ \frac{\partial B}{\partial z} + \frac{ iL}{2k}\nabla^2_\perp B = \frac{\alpha_lL}{2}(1+i\Delta) A(q,p)B \\
\end{array} \right.
\end{equation}
where we have scaled $z$ to the thickness $L$ of the medium, $\alpha_l$ is the linear absorption coefficient, $\Delta (= 2\delta/\Gamma)$ is the scaled detuning. For a two-level system, the linear absorption coefficient can be written as $\alpha_l = \alpha_0/(1+\Delta^2)$, where $\alpha_0$ is the on-resonance absorption, and $\alpha_0 L$ is the on-resonance optical density (OD), which is an important figure of merit for a cold-atom cloud (OD=210 for the cloud in \cite{Camara2015}, see caption to Fig. \ref{fig:setup}).

The function $A(p,q)$ describes the nonlinearity of the atomic susceptibility, as modeled by  (\ref{exacteqs}), by some approximation thereto, or some other model, including other optical systems with phase-independent interaction of counterpropagating beams \cite{Honda1996}.
By definition, $A(0,0) =1$, but $A(p,q) \ne A(q,p) $ in general, because of non-reciprocity due to standing-wave effects. The cubic model ($A(p,q) = 1-p -(1+h)q$) is the simplest example, explicitly non-reciprocal if $h \ne 0$.

 \begin{figure}
\centering%
\includegraphics[width=\columnwidth]{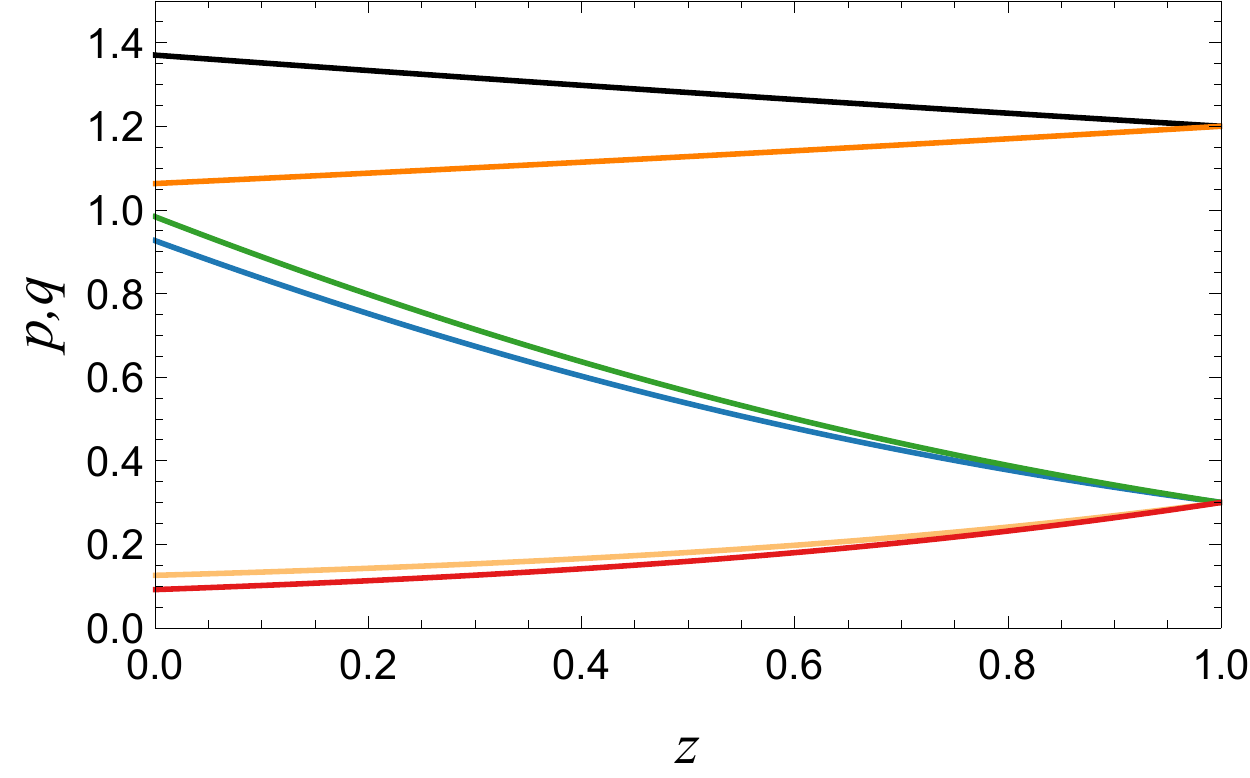}
 \caption{(Color online)
Dependence of zero-order intensities on the longitudinal coordinate $z$ scaled to the medium length~$L$, in a two-level medium with on-resonance optical density $OD=210$ (see Fig. \ref{fig:setup}): forward $p(z)$ and backward $q(z)$  for several cases. Lowest curves are for $\delta/ \Gamma = 5$, with output $p(1) =0.3$ and a $R=1$ mirror so that $q(1)=0.3$: upper and lower curves are for $h=0$, i.e. no reflection grating, inner curves for
 $h=1$. Uppermost curves are for larger detuning $\delta/ \Gamma = 10$ and $h=1$, to illustrate a case where absorption effects might be considered negligible, leading to a quasi-Kerr approximation to the two-level response.
}
\label{fig:zeroorder}
 \end{figure}

\section{Zero-order equations and solutions}
To find the pattern-formation thresholds, we first drop diffraction, and solve the plane-wave, zero-order problem in which $F,B$ depend on $z$ alone. From (\ref{transeqs}) it follows that the plane-wave intensities $p(z), q(z)$ obey the real equations:
\begin{equation}
\label{abzero}   \left \{
\begin{array}{l}
\frac{dp}{d z}  =-\alpha_l L A(p,q)p ,\\
\\ \frac{dq}{d z}  =\alpha_l L A(q,p)q \\
\\
\end{array} \right.
\end{equation}
leading to the expected exponential absorption of the intensities in the linear limit.

We define the input intensity $p(0)=p_0$ and transmitted intensity $p(1)=p_1$, and  similarly $q(0)=q_0$, $q(1)=q_1$. The boundary conditions of the SFM system are  $q_1 = R p_1$, where $R$ is the mirror reflection coefficient. 
We now solve (\ref{abzero}) for various two-level models.

For $h=0$, $A=1/(1+s)=1/(1+p+q)$ is symmetric in its arguments, and it follows that the product of the counter-propagating intensities (and indeed of the fields, $FB$) is independent of $z$, simplifying the analysis.  We set $p(z)q(z) = K$, where $K$ is constant, and thus  $K = p_1q_1 = Rp_1^2 $ for a feedback mirror of reflectivity $R$. It follows that the backward intensity $q(z)$ is given by $K/p(z)$, enabling the first equation of (\ref{abzero}) to be written in terms of $p(z)$ alone. It can then be integrated analytically, giving
\begin{equation}
\label{p_nograt}
\ln(p/p_0)+p-K/p - p_0 + K/p_0 +\alpha_lLz = 0 ,
\end{equation}
and hence, for the transmitted power $p_1$ (using the explicit SFM value of $K$):
\begin{equation}
\label{p_1nograt}   \\
\ln(p_1/p_0)+(1-R)p_1=p_0-Rp_1^2/p_0 -\alpha_lL .\\
\\
\end{equation}

The all-grating system given by (\ref{exacteqs}) also possesses a propagation constant for $h=1$, this time given by $K = W(z) - s(z)$, where $ W(z) = (1+2s+\xi^2)^{\frac{1}{2}}$, and $\xi(z) = p(z) - q(z)$. Essentially the same conservation law was noted by Van Wonderen et al in the context of optical bistability in a Fabry-Perot  resonator \cite{vanW1989}, for which the propagation equations are identical to the present case, though the boundary conditions are different.

In terms of $W, s,\xi$ the all-grating function $A_{all} (p,q)$ becomes
$A_{all} = (1+(\xi-1)/W)/(s+\xi)$, with its transpose $A_{all} (q,p)$ obtained by $\xi  \rightarrow -\xi$. Recasting equations (\ref{abzero}), the propagation equations for $s$ and $\xi$ take a fairly simple form:
\begin{equation}
\label{sandxi}   \left \{
\begin{array}{l}
\frac{ds}{d z}  = - \alpha_l L\xi/W ,\\
\\ \frac{d\xi}{d z}  = - \alpha_l L(1-1/W) \\
\\
\end{array} \right.
\end{equation}
from which one easily deduces $dW/dz =ds/dz$, and thus the constancy of $K = W(z) -s(z)$. One can then obtain an integrable differential equation in just one variable. For example, by using the definitions of $W$ and $K$ to express $W$ in terms of $K$ and $\xi$, the second of equations (\ref{sandxi}) is easily integrated to yield:
\begin{equation}
\label{xi_all}   \\
\xi + \ln(\xi + (\xi^2+2-2K)^{\frac{1}{2}}) +\alpha_l L z = const.\\
\\
\end{equation}
For the important case $R=1$, we have $s_1 = 2p_1$, $\xi_1 = 0$, hence $W_1 = (1+4p_1)^\frac{1}{2}$ and thus $K = (1+4p_1)^\frac{1}{2} -2p_1$. Using this data in (\ref{xi_all}) yields an implicit expression for $\xi_0$ in terms of $K$ (and thus $p_1$):
\begin{equation}
\label{xi_0}   \\
\xi_0 + \ln(\xi_0 + (\xi_0^2+2-2K)^{\frac{1}{2}})  -{\frac{1}{2}} \ln(2-2K) = \alpha_l L .\\
\\
\end{equation}
Given $\xi_0$, it is straightforward to calculate $W_0$ and $s_0$, and thus the input intensity $p_0$ and
the backward output intensity $q_0$, all in terms of the given transmitted intensity $p_1$, thus completing the solution of the plane-wave problem for the all-gratings model.

For the MM model $A(p,q)=(1+p)/(1+s)^2$. We can again find a propagation constant, in this case given by $K=pq/(1+s)$, again leading to a an integrable first-order equation in $p(z)$ alone. It turns out that the MM transmission shows ``bistability", i.e. the output $p_1$ is not a single-valued function of the input $p_0$, if  $\alpha_l L$ is big enough.

This is surprising and counterintuitive, and turns out to be a flaw in the model: including more terms in the series expansion (\ref{sigmas}) eventually makes $p_1$ single-valued. In particular the all-gratings formula (\ref{xi_all}) and its $R=1$ sub-case (\ref{xi_0}) give single-valued transmission characteristics. We therefore drop further detailed consideration of the MM model.


Figure \ref{fig:zeroorder} illustrates the $z$-dependence of the zero-order intensities in a two-level medium for several cases, with $OD=210$ as in the experiment illustrated in Fig. \ref{fig:setup}. The lowest group of curves are for moderately high absorption, $\alpha_l L \sim  2$, at $\delta/ \Gamma = 5$, and chosen to illustrate the two cases $h=0$  described by (\ref{p_nograt}) and $h=1$, where the $z$-dependence may be deduced from (\ref{xi_all}). To assist comparison, we assume the same output $p_1 =0.3$ and a perfect mirror so that $q_1=0.3$ also. The differences are fairly slight, the no-grating case having a slightly higher effective absorption for both forward and backward intensities. As we will see, there is a much more profound difference in the instability thresholds. We also display full-grating curves for larger detuning $\delta/ \Gamma = 10$, to illustrate a case where absorption effects might be considered negligible, leading to a quasi-Kerr approximation to the two-level response, which we will analyze below.


 \section{Transverse perturbations}
\label{sec:transpert}
We now assume that a solution has been found for the plane wave case: $F=F_0(z)$, $B=B_0(z)$, obeying appropriate longitudinal boundary conditions. This solution may be numerical, or a solution to some special-case or approximate version of  (\ref{transeqs}). We now turn our attention to the stability of such a plane wave solution against transverse perturbations.

We consider perturbations of the form $F=F_0(1+f\cos(Qx))$, $B=B_0(1+b\cos(Qx))$, where ($f,b$) are complex ($z$-dependent) amplitudes of the transverse mode function $\cos(Qx)$, chosen without loss of generality to respect the transverse symmetries of (\ref{transeqs}) and the mirror boundary conditions. The transverse perturbation has wave vector $Q$, corresponding to a diffraction angle $Q/k$ in the far field. We define a diffraction parameter  $\theta=Q^2L/2k$, physically the phase slippage between the $f$ and $F_0$ in traversing the cloud. Because $Q$ is experimentally a free parameter, so is $\theta$, and we have to calculate threshold intensities as a function of $\theta$, anticipating that the $Q$ corresponding to the lowest threshold will be dominant in any experiment, especially a pulsed experiment.

We assume that the fields $(f,b)$  are time-independent, adequate to calculate the threshold of a zero-frequency pattern-forming (Turing) instability at wavevector $Q$. To find Hopf instabilities, or to properly account for dynamical behavior of the field-atom system, we would have to start from the Maxwell-Bloch equations, rather than our susceptibility model. It is worth mentioning that van Wonderen and Suttorp, in a later paper on dispersive optical bistability \cite{vanWonderen91}, perform a perturbation analysis of the full Maxwell-Bloch equations with all grating orders included (though without transverse effects). The resulting model is very involved, and beyond our present scope. Meantime, we are content to address the Turing pattern threshold problem.

Within this constraint, we can say nothing about the nature and symmetry of the pattern which actually forms once threshold is exceeded. However, we know that hexagonal patterns are generic in systems of the type under consideration, and indeed are the dominant pattern observed in the experiments reported in \cite{Camara2015}. In a sense, therefore, threshold calculation is the most important step towards establishment of a theoretical underpinning for the observations of Camara et al \cite{Camara2015} and related experiments.
Assuming $|f|,|b| << 1$, we thus obtain the linearised propagation equations:

\begin{equation}
\label{thickpert}   \left \{
\begin{array}{l}
\frac{df}{dz} = - i\theta f  -\alpha_lL(1+i\Delta)(A_{11}f'+A_{12}b') ,\\
\frac{db}{dz} = i\theta b +  \alpha_lL (1+i\Delta)(A_{21}f'+A_{22}b') \\
\end{array} \right.
\end{equation}
Here $f = f' +i f''$, $b=b' + ib''$, and the real quantities  $A_{ij}$ are  defined as $A_{11}=p\frac{\partial A(p,q)}{\partial p}$,
$A_{12}=q\frac{\partial A(p,q)}{\partial q}$,
$A_{21}=p\frac{\partial A(q,p)}{\partial p}$,
$A_{22}=q\frac{\partial A(q,p)}{\partial q}$, and form a $2 \times 2$ matrix,  $\hat{A}$.

 In the presence of absorption, the elements of $\hat{A}$ are z-dependent, for example obeying the zero-order solutions derived above for  various models, and usually no analytic solution for $f(z),b(z)$ is available, requiring a resort to numerics. Below, we will consider both numerical investigations of the full (absorptive) model, as well as simpler models, including the quasi-Kerr case, in which the detuning is large enough to neglect the absorption, enabling analytic solution of the perturbation equations.

We have to solve (\ref{thickpert}) subject to appropriate boundary conditions. As there is no input field perturbation, we set $f_0 = f(0) = 0$. The counter-perturbation field at $z=0$, $b_0 =b(0)$, is physically determined by its value at $z=1$, but  the system  (\ref{thickpert}) is mathematically well-defined and solvable for any given $b_0$. Given initial conditions $(f,b)_{z=0}=(0, b_0)$, numerical integration of  (\ref{thickpert}), using the known functions $p(z) = |F_0|^2$ and $q(z)=|B_0|^2$, generates a pair of complex output perturbation fields at $z=1$, namely ($f_1,b_1$). For an acceptable solution, these fields must obey appropriate physical boundary conditions at $z=1$. For the SFM system these are  given by $f=b$ (note this is independent of mirror reflectivity $R$, because of the definition of ($f,b$) as relative perturbations).

\begin{figure}

\includegraphics[scale=0.9,width=\columnwidth]{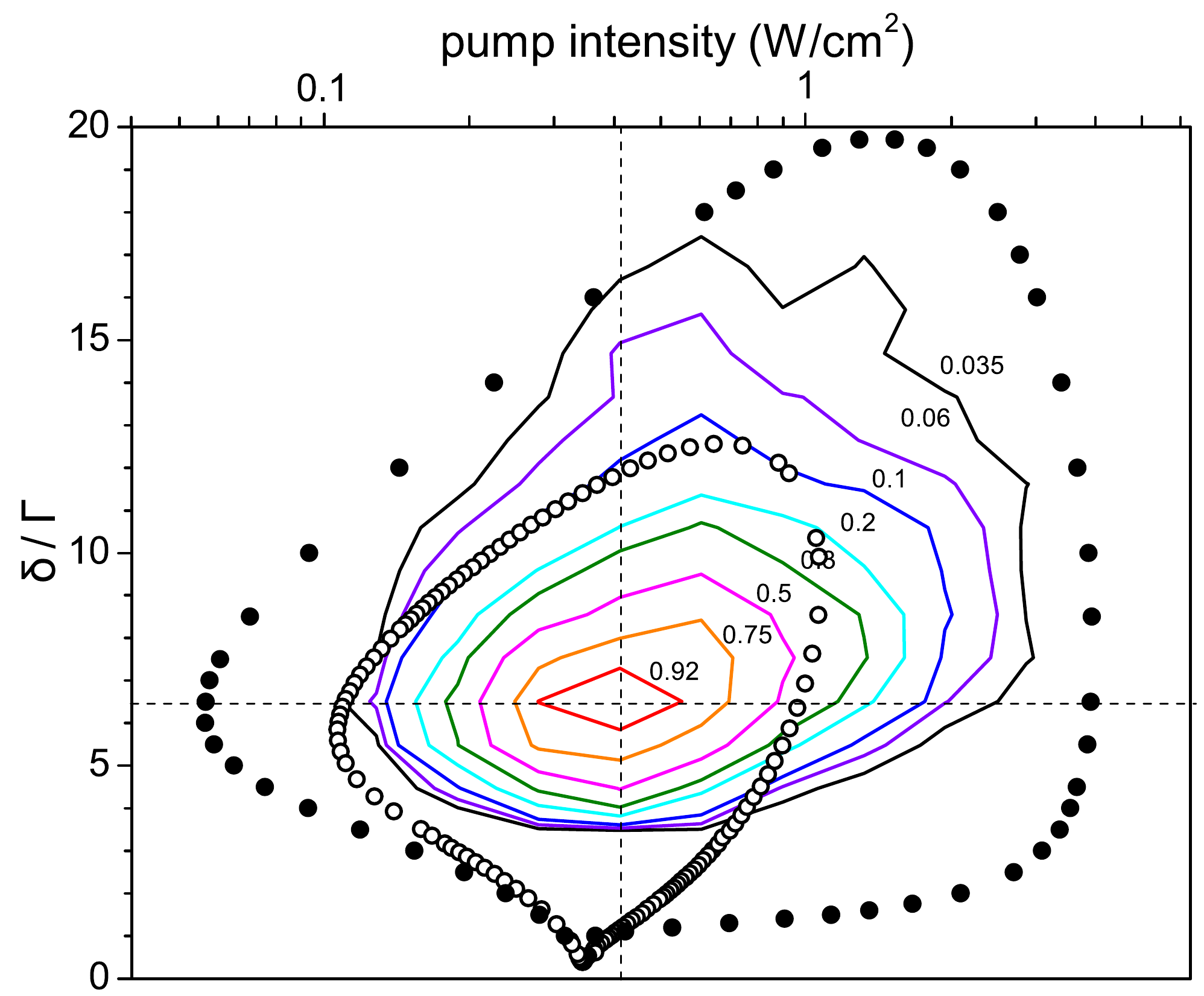}
\caption{ (Color online) Two-level instability domain ($\delta > 0$) reported in \cite{Camara2015}.
Diffracted power  $P_d$  is measured as a function of $\delta > 0$ (note that $\Delta=2\delta/\Gamma$) and input intensity $I$. Note the logarithmic horizontal scale.
The dotted loops indicate maximal instability domains calculated in the thin-medium approximation as described in  \cite{Camara2015}: (full circles) domain calculated from  (\ref{xi_0}), i.e. with all reflection gratings included ($h=1$);  (open circles) domain calculated from  (\ref{p_1nograt}), i.e. with no reflection gratings ($h=0$). Both dotted traces are rescaled to absolute values of intensity and detuning.}
 \label{CamaraTuning}
 \end{figure}


Turning now to the solution of  (\ref{thickpert}), the fact that $f$ has to grow through the medium makes it useful to define an output  ``gain" $g= f_1/b_1$. Since $f=b$ on the mirror of an SFM system, we immediately conclude that $|g| = 1$ is a necessary condition for SFM instability.
We can expect that $g \sim 0$ at low intensities, when the nonlinearity is negligible. As the intensity is increased, $f$ and $b$ begin to couple through the interaction matrix $\hat{A}$, and we can expect the gain to increase, leading to instability if the parameters permit. As mentioned, our present approach cannot describe behavior above threshold, but if the nonlinearity saturates, as is true for a two-level system, $|g|$ may begin to decrease for large enough input intensity. Then the system may re-stabilize, and the pattern will disappear. This scenario is illustrated in  Figure~\ref{CamaraTuning}, which  compares the threshold domains for two two-level absorptive models with  experimental data  \cite{Camara2015} on the detuning behavior of the diffracted power observed under pattern formation conditions in a cold Rb cloud with single feedback mirror. There is a minimum and maximum detuning for the observation of the SFM instability, while between these limits there is both a lower and an upper threshold power, with patterns  observed only at intermediate powers. The computed threshold loops in Figure~\ref{CamaraTuning} correspond to approximate ``thin-medium" models with and without short-period (reflection) gratings. The loop for the ``with" case is in much better agreement with the experimental results than that for for a similar model without such gratings, for which the loop is much smaller, and does not span the experimental domain. Note that the presence of reflection gratings has a much larger effect on the instability thresholds (about a factor of two) than on the zero-order intensities, where the effect is modest (Fig. \ref{fig:zeroorder}).

\section{Gain circle}
The transverse gain function $g=f_1/b_1$ is complex, and its phase as well as its magnitude must satisfy the boundary conditions at $z=1$, which depend on the mirror displacement. If the mirror displacement is $DL$ (Fig.~\ref{fig:setup}), then the boundary condition is $b_1 = e^{-2i\psi_D} f_1$, where $\psi_D= D \theta $. (Note that $D$ can be negative if the feedback optics involves a telescope.) Thus the complete boundary condition is that $g = e^{2i\psi_D}$, i.e. $g$ must lie at a point, the threshold point, on the unit circle in the complex plane.

Before looking at specific examples, there are some general considerations which give insight into methodology, but also into the physics. Because  (\ref{thickpert}) is a linear system, its solutions obey the principle of superposition. Hence, if input condition $(f_0,b_0) = (0,1)$ generates outputs $(f_1,b_1) = (f_r,b_r)$ and input condition $(f_0,b_0) = (0,i)$ generates outputs $(f_1,b_1) = (f_i,b_i)$, then an arbitrary  input condition $(f_0,b_0) = (0,u+iv)$, with $(u,v)$ real, generates outputs $(f_1,b_1) = (uf_r+vf_i,ub_r+vb_i)$. The gain is then given by $g=g(u,v)=(uf_r+vf_i)/(ub_r+vb_i)$. Thus, for any given physical parameters, one need only obtain the pairs $(f_r,b_r)$ and $(f_i,b_i)$, and then testing for the SFM instability is a matter of algebra.

In looking for a solution, a graphical approach is convenient and instructive. Some algebra shows that the points of the gain function $g(u,v)$ always belong to a circle.  This ``gain circle" is given by a simple analytic formula  in terms of $g_r =g(1,0)=f_r/b_r$ and $g_i =g(0,1)=f_i/b_i$:
\begin{equation}
\label{gaincircle}
g(\phi)=g_i +(g_r - g_i)(1-e^{2i\phi})/( 1-e^{2i\phi_0})
\end{equation}
where $\phi$ is a free parameter which traces out the gain circle, while $\phi_0$ is the phase of $b_i/b_r$.

For finite $\theta$ the phase of the threshold point, the feedback phase, will vary as $D$ is varied, causing the threshold point to trace out all or part of the unit circle. Hence the intersections, if any, of the gain circle with the unit circle define instability thresholds for the mirror displacement(s) $D$ corresponding to the  intersection(s).

%
 \begin{figure}
\centering%
\includegraphics[width=\columnwidth]{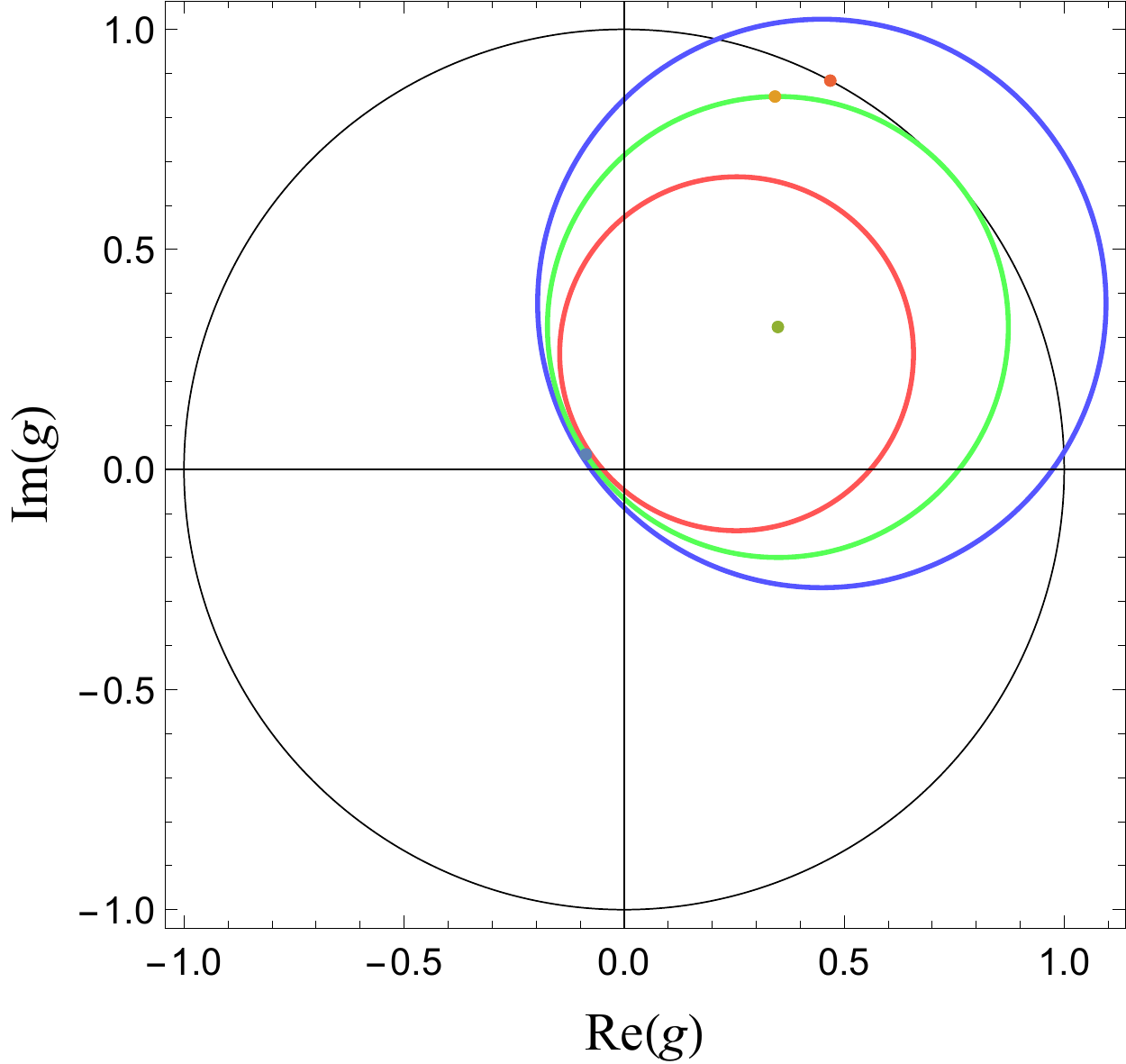}
 \caption{ (Color online) Illustration of transverse gain circles (see text) calculated from  (\ref{thickpert}) for different input intensities. The parameters here are: $OD=210$, $\delta/\Gamma=2$, $\theta=2$. The unit circle centred on the origin is the locus of the feedback phase as mirror displacement $D$ is varied. Lying on it, the dot (red online) is the feedback phase for the particular case $D= -1.3$. The displaced circles are the loci of transverse gain for three cases: (a) the smallest gain circle (red online) lies wholly inside the unit circle, and so the system is always below threshold for this case (scaled input intensity $p_0 =7.90564$); (b) the middle gain circle (green online) touches the unit circle, and so the system reaches threshold for one value of $D$ (scaled input intensity $p_0 =8.1266$); (c) the largest gain circle (blue online) intersects the unit circle at two well-spaced points, and so the system is above threshold for a wide range of $D$ values, including $D= -1.3$ (scaled input intensity $p_0 =8.29754$). Points on the arc of the touching circle corresponding to $g_r$ (blue online) and $g_i$ (brown online) are also shown. Its center is also marked with a (green online) dot.
 }
 \label{fig:circles}
 \end{figure}


Figure \ref{fig:circles} illustrates  typical cases for system (\ref{thickpert}). 
As expected, the gain circle lies wholly within the unit circle when the input intensity is low, so that there are no intersections, and thus no instability. At higher intensity, the gain circle intersects the unit circle at two points, and there is instability for all mirror displacements $D$ for which the feedback phase lies on the arc between the two intersections for which the gain circle lies outside the unit circle. Because the feedback phase  $ e^{2i\psi_D} $ is periodic in $D$, such thresholds are periodic in mirror displacement, with a period which depends on $Q$ through $\theta$. This is an example of the Talbot effect,
 whereby a transversely-periodic light field self-reconstructs under propagation through multiples of the Talbot period, $z_T=4\pi k/Q^2$ \cite{talbot36,ciaramella93}. Such $D$-periodicity of instability thresholds is observed experimentally, and will be discussed in more detail below.

An interesting and important intermediate case illustrated in Figure \ref{fig:circles} occurs when the gain circle touches the unit circle. This corresponds to the lowest possible threshold for any $D$ at these parameters (modulo Talbot recurrences). This minimum threshold will be achieved for some value of $D$ if it is varied over a Talbot period.
The implication is that the locus (or loci) in the ($\theta,p_0$) plane of tangencies between the gain circle and the unit circle forms an envelope curve (or curves) bounding the set of threshold curves in the ($\theta,p_0$) plane corresponding to any set of $D$ values. Given the analytic formula (\ref{gaincircle}) for the gain circle, it is straightforward to find $(\theta,p_0)$ pairs such that the gain circle touches the unit circle, thus tracing out envelope curves in the $(\theta,p_0)$ plane. It is similarly straightforward to find $p_0$ and $\theta$ such that the gain circle intersects the unit circle at the feedback phase  corresponding to any given mirror displacement $D$, and thus to trace out threshold curves for that $D$. Examples, and implications, of envelope and threshold curves for various models will be presented below.


\section{ Two-level System Envelopes and Thresholds}

\begin{figure}

\includegraphics[scale=0.9,width=\columnwidth]{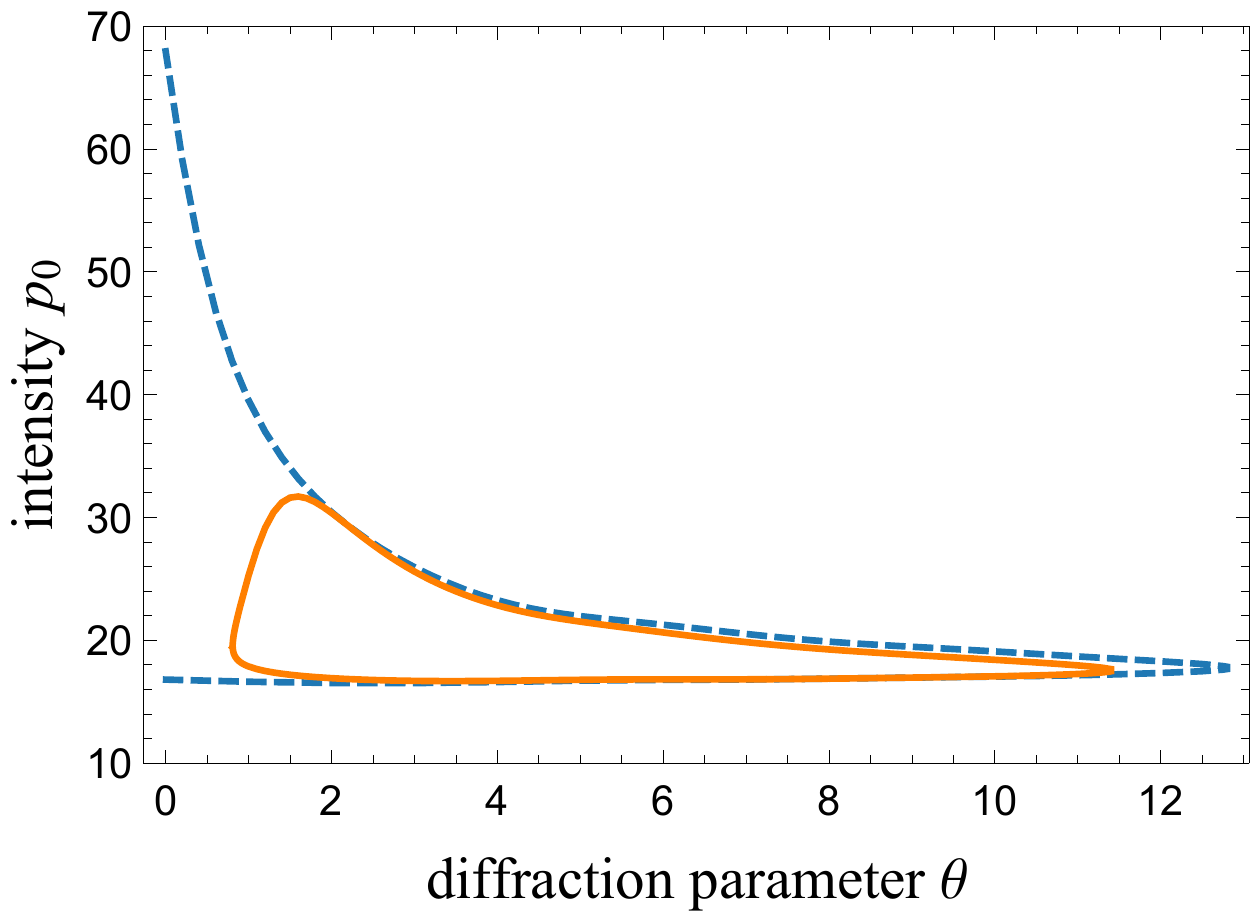}
\caption{ (Color online) Threshold and envelope curves calculated from (\ref{thickpert}) for a two level system with all gratings included ($h=1$) with $R=1$ feedback mirror. Scaled input intensity  $p_0$ is plotted against diffraction parameter $\theta=Q^2L/2k$. Outer (blue online) curve is the envelope curve, the limiting threshold for any mirror displacement: inner (orange online) is the threshold curve for mirror displacement $D=0$, which, close to its maximum, touches the envelope curve. It also touches the envelope at low values of $p_0$, in fact almost coinciding with the envelope curve over a wide range of $\theta$. The envelope curve has finite intercepts with $\theta=0$ axis (see text for discussion).
Other parameters: $OD=210$, $\delta/\Gamma=1.5$. }
 \label{EnvD0del1p5}
 \end{figure}

As a first detailed example, we consider the two level system to be fairly close to resonance, with blue detuning   $\delta/\Gamma=1.5$. For optical density OD=210 (Fig.~\ref{CamaraTuning}) this corresponds to
$\alpha_l L= 21$, i.e. the linear absorption is very high. Such conditions have not been modeled before, except in thin-medium or no-grating approximations. Figure \ref{EnvD0del1p5} shows the envelope curve for this case,  together with the threshold for the mirror displacement $D=0$, calculated using the gain circle technique. As might be expected, the minimum threshold is rather high, $p_0 \sim 17$, which means that substantial saturation is required - the output intensity $p_1$ is of order unity in the low-threshold region. There is also an upper threshold, essentially the bleaching of the absorption destroys the nonlinearity. Here $p_1$ is of the same order as $p_0$. As predicted, the threshold curve lies inside the envelope curve, touching it at closest approach.

Whereas the $D=0$ threshold curve avoids $\theta=0$, which is typical behavior for SFM models, the envelope seems to have finite intercepts at $\theta=0$. To interpret this, we note that the feedback phase $\theta D$ tends to zero as $\theta \to 0$ for any finite $D$. Thus the corresponding threshold point gets trapped close to the positive real axis, away from the envelope-defining contact between the gain circle and the unit circle, which will generally occur at a finite phase angle. If we also allow $D$ to increase without limit, however, finite feedback phase, and hence finite thresholds, can be sustained as $\theta \to 0$. Now the ``thin-medium" approximation, in which the diffraction within the medium is considered negligible compared to that in the feedback loop, implies $D \sim d/L$ diverges. Thus we identify the intercept of the envelope with the $\theta$ axis as exactly the thin medium limit. Indeed, this is confirmed for our case. The intercepts of the envelope found using the gain circle technique coincide exactly with those we calculated previously by direct use of the thin-medium approach, and the results of which were presented in \cite{Camara2015}. We will return to this issue below, when we consider other models.

Another question arising from the finite intercept of the envelope curve is how to interpret its continuation to negative $\theta$, which presents no numerical difficulties (for diffractively thin media negative feedback distances were first considered in \cite{ciaramella93}). If we look at the structure of (\ref{thickpert}), we observe that simultaneously changing the sign of $\theta$ and $\Delta$ has the effect of transforming the equations into their complex conjugates. The boundary condition is also conjugated. Thus we can interpret the continuation of the envelope curve(s) to negative $\theta$ as corresponding to the opposite sign of detuning. We will routinely take advantage of this symmetry to present result for both signs of detuning in a single diagram.
An important corollary is that SFM thresholds are equal for both signs of detuning in the thin-medium limit for all models described by (\ref{thickpert}). In contrast, the finite slope of the envelope curve in
Fig. \ref{EnvD0del1p5}  at its intercepts with $\theta =0$ implies that there is no red-blue symmetry when diffraction in the medium is taken into account.

\begin{figure}

\includegraphics[scale=0.9,width=\columnwidth]{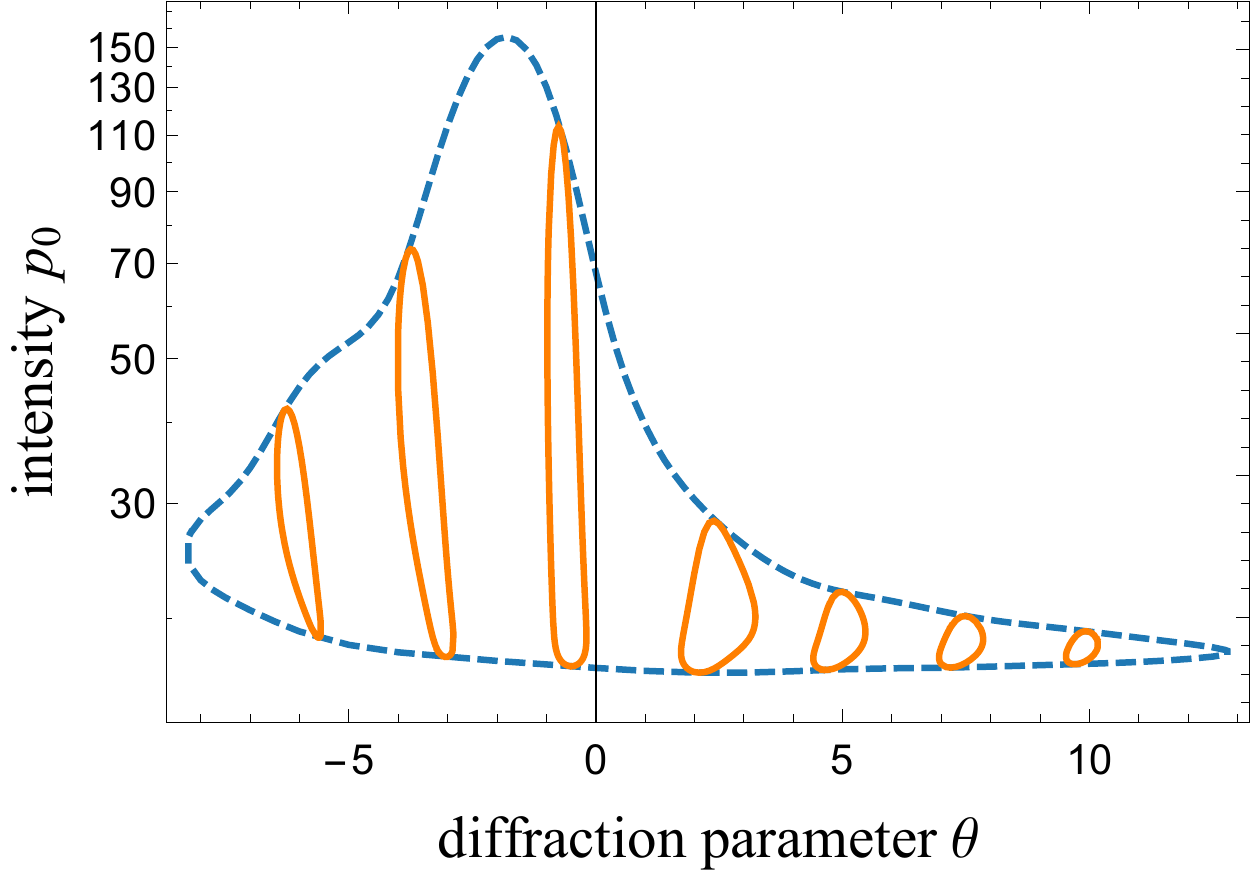}
\caption{ (Color online) Threshold and envelope curves calculated from (\ref{thickpert}) for the same conditions as Fig. \ref{EnvD0del1p5}, except that the feedback mirror displacement is $D= -1.3$, which corresponds to the experimental results of Fig. \ref{CamaraTuning}.  Scaled input intensity  $p_0$ is plotted (here on a log scale, for clarity) against diffraction parameter $\theta=Q^2L/2k$, which is continued to negative $\theta$ (see text) so as to present results for red, as well as blue, atomic tuning. The envelope curve, the continuation to negative $\theta$ of that in Fig. \ref{EnvD0del1p5}, shows a large red-blue tuning asymmetry. Inside the envelope is a  set of discrete closed threshold loops for $D= -1.3$, each of which touches the envelope above and below.
 }
 \label{EnvDnegdel1p5}
 \end{figure}

 In Fig. \ref{EnvDnegdel1p5} we use this tuning-diffraction correspondence to extend the envelope, and also to display threshold curves for mirror displacement $D=-1.3$, which corresponds to the experimental results of Fig. \ref{CamaraTuning}. The extended envelope displays a huge red-blue tuning asymmetry in the upper threshold, and a smaller one in the lower threshold, for which blue tuning gives the lowest thresholds, in accord with experimental experience. The threshold curves for fixed $D=-1.3$ are very different from that for $D=0$ in Fig. \ref{EnvD0del1p5}, being a discrete set of closed loops, which each touch the envelope twice,  close to their upper and lower extrema.

\begin{figure}

\includegraphics[scale=0.9,width=\columnwidth]{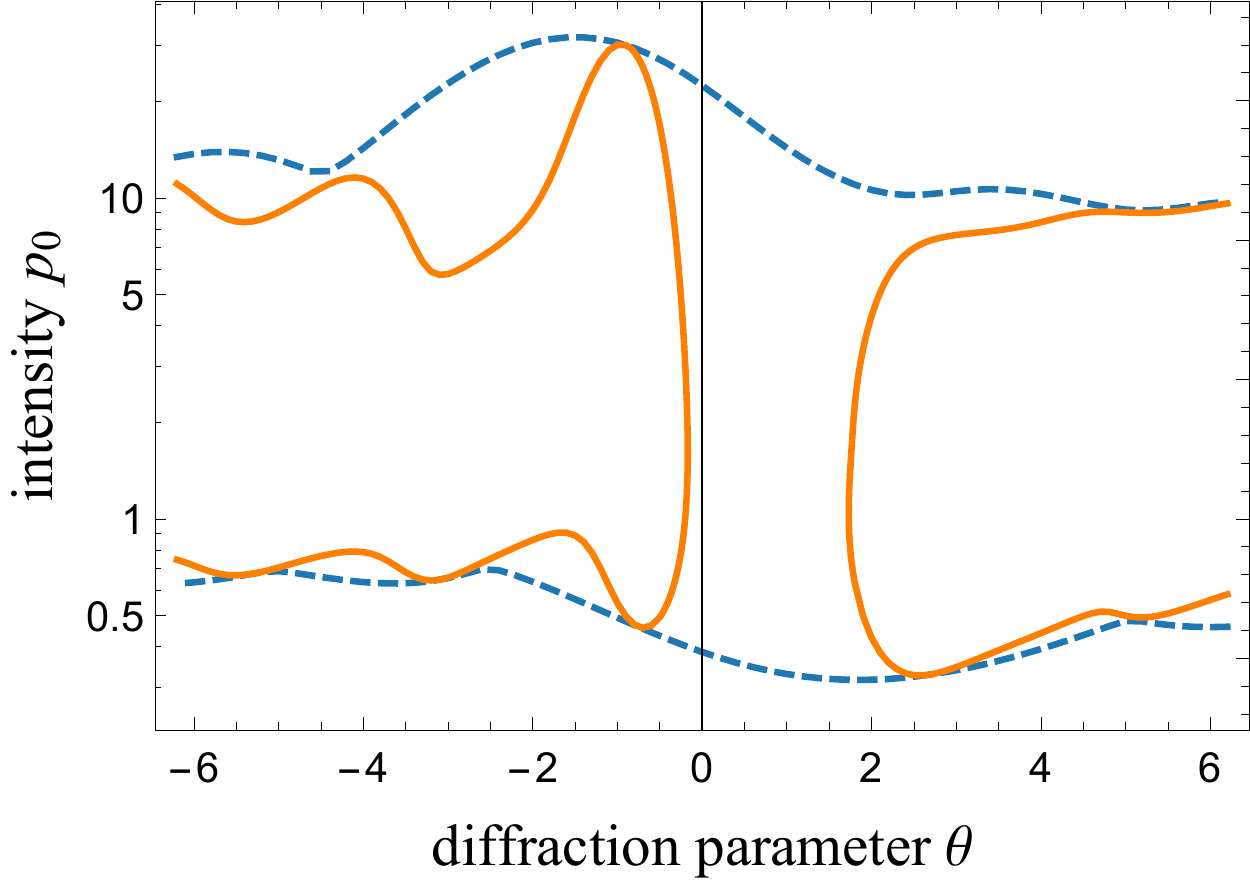}

\caption{ (Color online) Threshold and envelope curves calculated from (\ref{thickpert}) for the same conditions as Fig. \ref{EnvDnegdel1p5}, except  $\delta/\Gamma=5$.  Scaled input intensity  $p_0$ is plotted (again on a log scale, for clarity) against diffraction parameter $\theta=Q^2L/2k$, which is continued to negative $\theta$ (see text) so as to present results for red, as well as blue, atomic tuning. Upper and lower portions of both envelope and threshold curves are well separated for large $\theta$, asymptotically corresponding to phase-conjugate oscillation thresholds. }
 \label{Thresholds}
 \end{figure}

Increasing the magnitude of the detuning, both the absorptive and the dispersive  nonlinearity decrease, but at different rates, with the absorption decreasing faster, which favors pattern formation. Fig. \ref{CamaraTuning} shows that the pattern threshold intensity is a minimum, and its intensity range a maximum, for detunings of magnitude $\sim 5$. Figure \ref{Thresholds} illustrates  envelope curves, and threshold curves for $D=-1.3$, vs diffraction parameter for $\delta/\Gamma=5$, with other parameters as before. For this case, both the envelope and the fixed-D threshold curves seem to be open to large $|\theta|$, indicating that low (but not lowest) thresholds persist to large diffraction angles (divergent $Q$). This is not unexpected, because the coupling of the $f$ and $b^{*}$ components of the transverse perturbations is phase-conjugate (PC) in nature, and so is phase-matched for all diffraction angles. As was discussed for counter-propagation in  Kerr media by Firth et al \cite{firth90b}, at small diffraction angles the non-phasematched couplings of $f$ and $f^*$, and of $f$ and $b$ (and analogously for $b$'s couplings) give additional oscillatory contributions to the transverse gain, and can lead to thresholds which are significantly below the PC oscillation threshold \cite{grynberg93}. Similar considerations apply in our case, though the SFM boundary conditions and the two-level nonlinearity lead to quantitative differences.

Fig. \ref{Thresholds} displays oscillations in both the envelope and the threshold curves, for both signs of detuning, though more prominent for red-detuning. The $D=-1.3$ threshold curves are again wholly contained by the envelope curves, with touching contact at several points. There are several near-contacts, linked to the complexity of the system in such strongly-nonlinear regions. The minimum and maximum thresholds are associated with tangencies in all case, however.

\begin{figure}

\includegraphics[scale=0.9,width=\columnwidth]{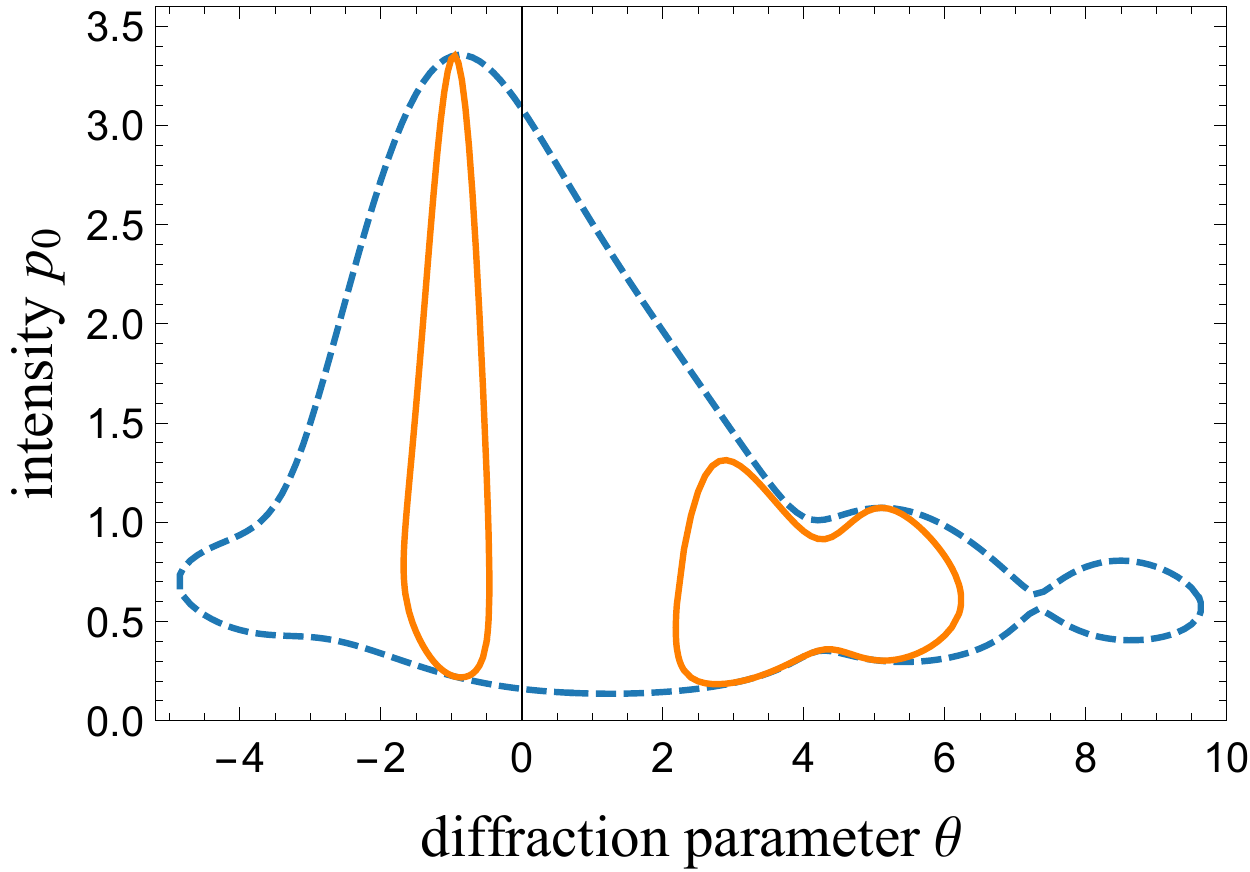}
\caption{ (Color online)  Threshold and envelope curves calculated from (\ref{thickpert}) for the same conditions as Fig. \ref{EnvDnegdel1p5}, except  $\delta/\Gamma=13.1$. Note that $p_0$ is here plotted on a linear scale.}
 \label{LargeDelta}
 \end{figure}

Further increasing the detuning leads to a fall-off in nonlinearity, and the envelopes begin to close again,  PC oscillation becomes impossible, and eventually the SFM transverse instability also disappears, at a detuning which depends on optical density OD. Fig. \ref{LargeDelta} shows the onset of this process, for detuning
$\delta/\Gamma=13.1$, other parameters as in the previous figures.  At such large detunings, absorption becomes small, and it is of interest to compare Fig. \ref{LargeDelta} with the corresponding results in the quasi-Kerr case (discussed in the next section), in which absorption is neglected, enabling analytic solution to the thus simplified version of system  (\ref{thickpert}).

\begin{figure}

\includegraphics[scale=0.9,width=\columnwidth]{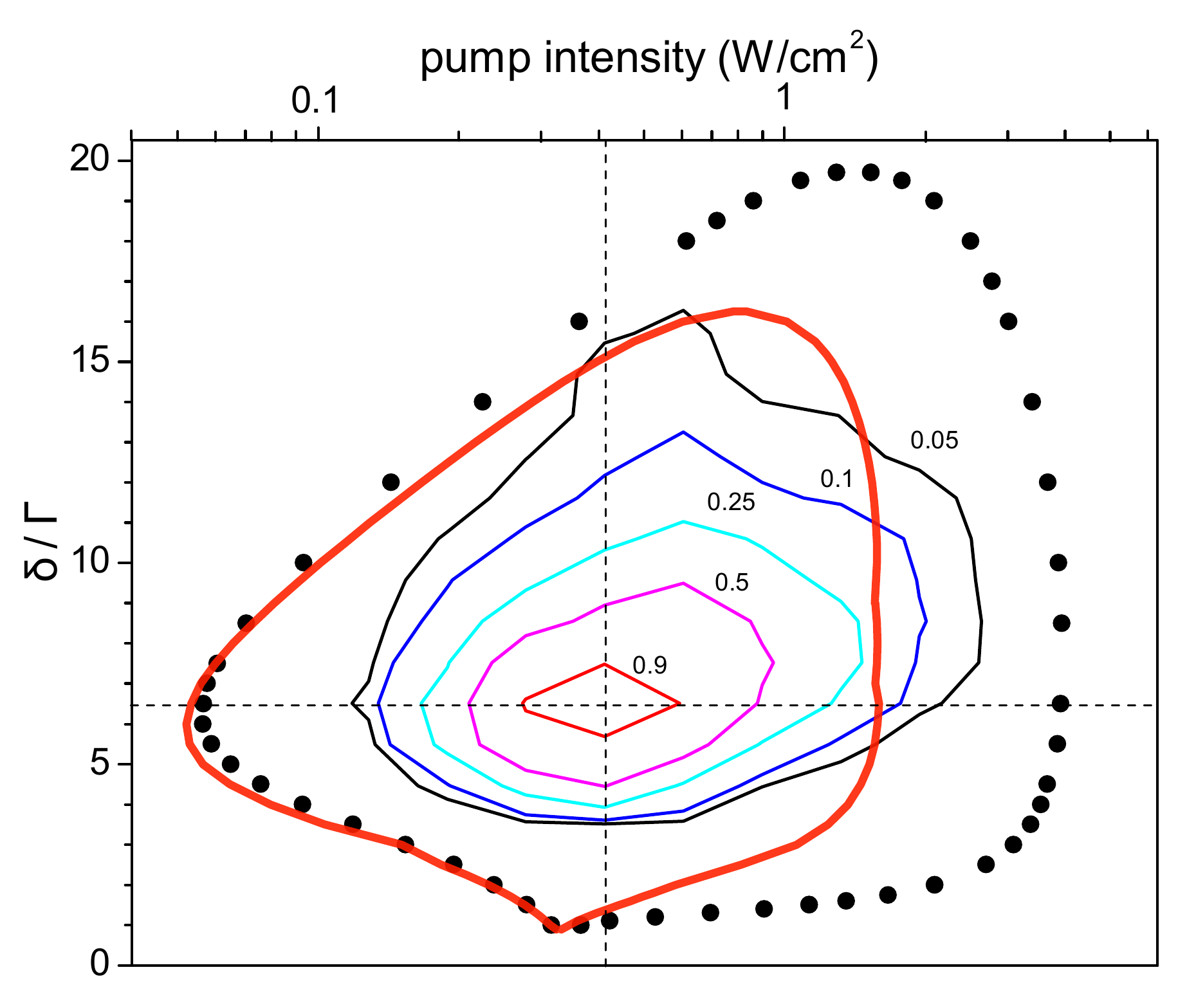}
\caption{ (Color online) Two-level instability domain, range of threshold input intensity $p_0$ in terms of  $\delta/\Gamma$ , with a logarithmic horizontal scale, as in Fig. \ref {CamaraTuning}. The larger loop (dots, black online) is as presented in \cite{Camara2015}, calculated in the thin-medium approximation for $R=1$, but identifiable as the $\theta=0$ envelope (see text), the smaller (red online) is calculated from (\ref{thickpert}), i.e. with all reflection gratings included ($h=1$), and for  $R=0.95$ and $D=-1.3$. $OD=210$. The contour plot loops show experimental data of Fig. \ref{CamaraTuning}, for comparison.}
 \label{AbsLoops}
 \end{figure}

Similar threshold calculations enable the minimum and maximum  thresholds to be found over the full range of detuning for which instability exists for a given configuration. Choosing parameters $D=-1.3$ and $R=0.95$ to align with the recent experiment  \cite{Camara2015}, we have calculated the instability domain using
the above methods based on the full thick-medium model (\ref{thickpert}). Results are shown in Fig. \ref{AbsLoops}. The instability domain is broadly similar to that found for the thin-slice model used in Fig. \ref{CamaraTuning}, though with a significantly smaller upper threshold. As mentioned above, the thin-medium threshold corresponds precisely to the $\theta=0$ intercepts of the envelope curves. In all the tuning cases shown, Figs. \ref{EnvDnegdel1p5},  \ref{Thresholds}, \ref{LargeDelta}, the upper intercept is substantially above the highest upper threshold for fixed $D=-1.3$, and Fig. \ref{AbsLoops} shows this to be the case for all tunings. The lower threshold, which is perhaps the most interesting experimentally, is very similar for both thin-medium and fixed-$D$ cases.

The agreement with experiment of the all-grating models is rather satisfactory, bearing in mind that the theory only calculates threshold conditions, while the experiment detects diffracted power only if the perturbation gain is large enough to build a strong pattern from noise within the microsecond or so duration of the pump pulse. Moreover, we note that the no-grating threshold domain in Fig. \ref {CamaraTuning} is smaller than that in which transverse structure is observed. This provides firm evidence that reflection gratings are present in the cold-atom cloud, in agreement with expectations based on the inability of transport mechanisms to wash out susceptibility gratings at such low temperatures when such short input pulses are used.

The comparison between experimental and theoretical curves is further complicated by the fact that the theory uses an uniform plane wave and the experiment a Gaussian input beam. The Fourier transform to extract the power in the modulation was performed over an area with diameter equal to the beam waist radius (i.e.\ at the 60$\%$ power point). The pump power reported in Fig. \ref {AbsLoops} is the peak power. As a certain area of a least two pattern periods need to cross threshold for a sizeable effect, it is understandable that the experimentally detected threshold is higher than the predicted one. At the high intensity threshold, the center of the beam will become stable again but modulation still exists in the wings. Hence it makes sense that the plane-wave instability closes before the experimentally obtained threshold.

\section{Quasi-Kerr case}
While the above technique based on the gain circle is general and flexible, it yields little in the way of analytic insight in cases where strong nonlinear absorption leads to large and complicated changes in the forward and backward intensities in propagation through the medium. If we restrict to large enough detuning that the absorption can be considered negligible, however, it follows that $p$ and $q$ are constant in the medium, and analytic soluton to this ``quasi-Kerr" approximation to the thick-medium model (\ref{thickpert}) is possible. Formally, in such a  model, we suppose that $|\Delta|$ is large enough that $\alpha_l L$ can be neglected, but with $\alpha_l \Delta L$ finite, so that the nonlinearity is purely refractive, as is the case for a true Kerr medium, in which the refractive index changes linearly with intensity.

In the quasi-Kerr approximation the matrix $\hat{A}$  has constant coefficients, and the equations (\ref{thickpert}) become

\begin{equation}
\label{quasiKerr}   \left \{
\begin{array}{l}
\frac{df}{dz} = - i\theta f  - i \alpha_lL \Delta(A_{11}f'+A_{12}b') ,\\
\frac{db}{dz} = i\theta b + i \alpha_lL \Delta(A_{21}f'+A_{22}b') \\
\end{array} \right.
\end{equation}

Evidently the combination $\alpha_lL \Delta$ is an important strength parameter for the nonlinearity. Bearing in mind that $\alpha_l = \alpha_0/(1+\Delta^2)$, with $\Delta$ large by assumption, there is an obvious trade-off between nonlinearity and absorption.  We will proceed by solving (\ref{quasiKerr}), analytically where possible, and testing against the results derived above for the ``full" two-level model with absorption.

For feedback mirror boundary conditions, we have $q = R p$.
For the symmetric equal intensity case ($q=p$), $A_{11}=A_{22}=A_{sym}$ and $A_{12}=A_{21}= GA_{sym}$. The matrix $\hat{A}$ then has a simple symmetric form

\begin{eqnarray}
\label{Asym}
& & \hat{A}_{sym}= A_{sym}  \left(
\begin{array}{cccc}
 1 & G \\
G & 1
\end{array}
\right). \nonumber
\end{eqnarray}
 Both $A_{sym}$ and  $G$ are in general functions of $s=2p$, but are independent of $z$. \\

We now define $\psi_{1,2}^2 = \theta (\theta +\kappa \phi_{1,2})$, where the effective Kerr coefficient
$\kappa = \alpha_l L \Delta$.
($\phi_1, \phi_2$) are the eigenvalues of $\hat{A}$,
chosen such that
($\phi_1, \phi_2) \to A_{sym}(1-G,1+G)$ (the eigenvalues of $ \hat{A}_{sym}$)  as $q \to p$. Thus defined $\psi_{1,2}$  coincide exactly with the quantities $\psi_{1,2}$ used in \cite{firth90b,Geddes1994} in analyzing the Kerr CP case. It follows that the analysis developed in these papers for the symmetrically-pumped CP Kerr problem extends to the present quasi-Kerr case, in which both the strength of the nonlinearity and of the grating-coupling $G$ can be intensity dependent. 
Detailed consideration of the CP problem for a two-level system is a subject for future work.

We now present  explicit forms of the matrix $\hat{A}$ for various models of interest here. For the Kerr case, we have
\begin{equation}
\label{AKerr}
\hat{A}_{Kerr}= -
\left(
\begin{array}{cccc}
 p & (1+h)q \\
(1+h)p & q
\end{array}
\right).
\end{equation}
For $p=q$ this leads to $A_{sym} = -p$ and $G=1+h$ as expected.

For the MM model, we obtain
\\
\begin{eqnarray}
\label{Amur}
& & \hat{A}_{MM}= - \frac{1}{(1+s)^3}\\
& & \left(
\begin{array}{cccc}
 p(1+s)-2hpq & (1+h)q(1+s) -2hq^2 \\
(1+h)p(1+s)-2hp^2 & q(1+s) -2hpq
\end{array}
\right). \nonumber
\end{eqnarray}
For $p=q=s/2$ and $h=1$ the above expression for $\hat{A}_{MM}$ leads to $A_{sym} = - \frac{p}{(1+s)^3}$, while we find an intensity-dependent grating factor  $G=2+s$. This differs from the results of \cite{Muradyan2005}, wherein the given formulae imply  $G=2$.

The general function $A$ given in (\ref{exacteqs}) also leads to explicit expressions for the matrix $\hat{A}_{all}$.  In the absence of grating terms, i.e. for $h=0$, it simplifies to
\begin{eqnarray}
\label{Anogr}
& & \hat{A}_{h=0}= - \frac{1}{(1+s)^2}  \left(
\begin{array}{cccc}
 p & q \\
p & q
\end{array}
\right) \nonumber
\end{eqnarray}\\
 which leads to $A_{sym} = - \frac{p}{(1+s)^2}$. $G=1$, as expected, implying a zero eigenvalue for $\hat{A}_{h=0}$, and hence $\psi_1 = \theta$. (The MM model gives identical results for $h=0$.) \\

 With all  grating terms included, i.e. for $h=1$, we obtain

\begin{eqnarray}
\label{Aall}
& & \hat{A}_{all}=  \left(
\begin{array}{cccc}
(1+s)/W^3 - A & -2q/W^3  \\
-2p/W^3  &(1+s)/W^3  -A^T
\end{array}
\right)
\end{eqnarray}\\
where $A^T(p,q)= A(q,p)$. For equal intensities $W=\sqrt{1+2s }$ and $\xi =0$. Some calculation then shows that $G$ is approximately $2+2s$ for small $s$. For larger $s$, however, there is a strong departure from Kerr-like behavior, in that $A_{11}$ changes sign at $s= 1+ \sqrt{2}$, and it follows that $G$ is negative for higher values of $s$.\\

 Using analysis analogous to that  in  \cite{firth90b,Geddes1994}, but with SFM boundary conditions $f_0=0$, $b_1 = \exp{-2i\psi_{D}} f_1$, we obtain, for perfect mirror reflection ($R=1$),  the SFM  threshold condition \\
\begin{equation}
\label{2LSfbm}
c_1 c_2 +\left(\frac{\psi_2}{\psi_1}c_D^2+\frac{\psi_1}{\psi_2} s_D^2 \right)s_1 s_2 = c_D s_D\left( \beta_1 s_1 c_2 -\beta_2 s_2 c_1\right).
\end{equation}

Here $c_i=\cos\psi_i$; $s_i=\sin\psi_i$: $c_D=\cos\psi_D$; $s_D=\sin\psi_D$, and
$\beta_n = \left(\frac{\psi_n}{\theta}-\frac{\theta}{\psi_n}\right)$.

In the quasi-Kerr case the envelope condition whereby the gain circle in diagrams like Fig. \ref{fig:circles} touches the unit circle corresponds to  transition between complex and real $\psi_D$ as roots of (\ref{2LSfbm}).  This leads to the following envelope condtion:

\begin{equation}
\label{env}
4(c_1 c_2 +\frac{\psi_1}{\psi_2} s_1 s_2)(c_1 c_2 +\frac{\psi_2}{\psi_1} s_1 s_2) = ( \beta_1 s_1 c_2 -\beta_2 s_2 c_1)^2.
\end{equation}

\begin{figure}
 \includegraphics[scale=0.9,width=\columnwidth]{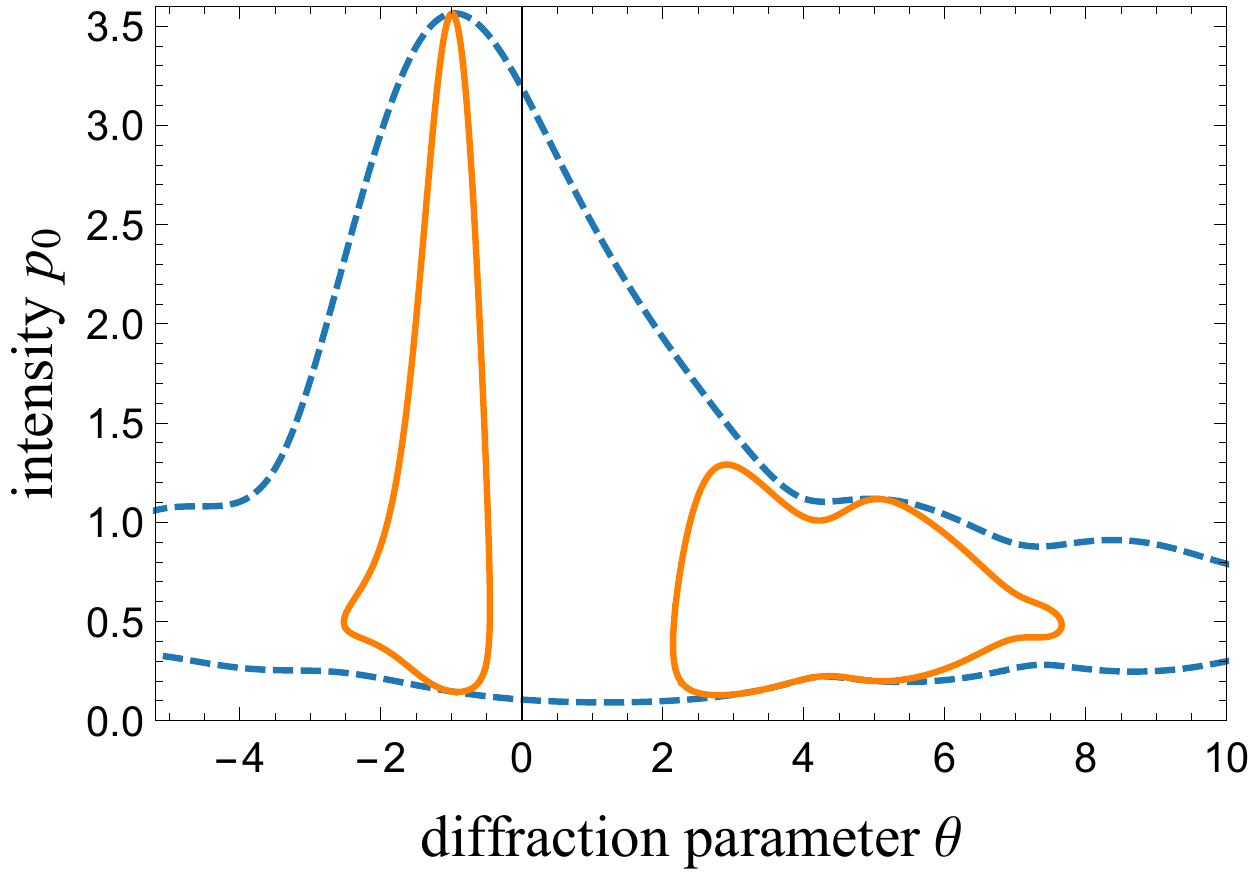}
\caption{(Color online)  Threshold and envelope curves. Blue curves (dashed): Envelope curves  calculated from (\ref{env}) for a two-level medium  described by $\hat{A}_{all}$, with $h=1$. Quasi-Kerr coefficient $|\alpha_l L \Delta| = 8$ and detuning $\delta/\Gamma=13.1$. Orange curves (solid): Threshold curves with a feedback mirror at negative effective distance ($D=-1.3$) from the end of the medium, which touches the envelope curves.
}

 \label{figQK7pt1}
 \end{figure}

As an example, Fig. \ref{figQK7pt1} illustrates envelope and threshold curves for the all-grating quasi-Kerr model, for a fairly small  quasi-Kerr coefficient,  $|\alpha_l L \Delta| =8$. There is a very good correspondence to the full model for the same parameters  (Fig. \ref{LargeDelta}). The main difference is that removing the small absorption losses make the instability and envelope domains slightly larger for the quasi-Kerr model. In particular, the range of $\theta$ is larger, extending to $ \sim 40$, but still finite, so that there is no phase-conjugate instability. 

\begin{figure}

\includegraphics[scale=0.9,width=\columnwidth]{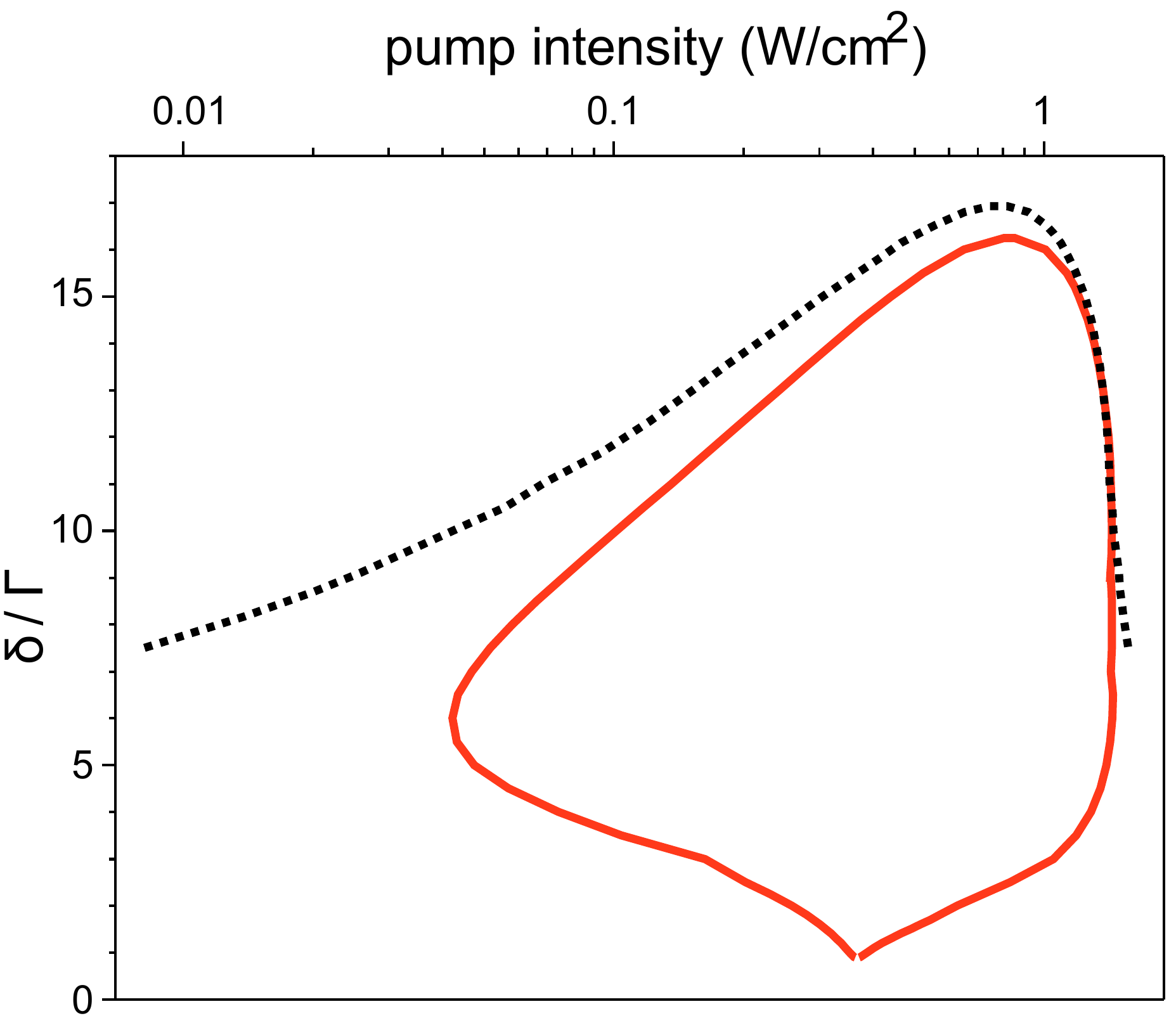}
\caption{ (Color online) Two-level instability domain, range of threshold input intensity in terms of  $\delta/\Gamma$, with a logarithmic horizontal scale. The closed loop (red online) is that calculated from (\ref{thickpert}) with absorption and all reflection gratings included ($h=1$). $R=0.95$, $D=-1.3$ and $OD=210$,  as in Fig.\ref {AbsLoops}. The open curve (dotted, black online) is calculated for the same parameters from   (\ref{2LSfbm}), as derived from the quasi-Kerr model equations (\ref{quasiKerr}). The latter curve is calculated only for $\delta/\Gamma > 7.5 $, because the  neglect of absorption in  (\ref{quasiKerr}) is untenable at small detunings.}
 \label{exvsqK}
 \end{figure}

A  key question is how useful the quasi-Kerr approximation is. To test this, we compare quasi-Kerr and ``exact" two-level thresholds over a range of tunings with other parameters equal, except that $R=1$ for the quasi-Kerr. Fig. \ref{exvsqK} shows such a comparison. Unsurprisingly, the fit is best at large detunings, with the quasi-Kerr model predicting lower thresholds which are increasingly underestimated as the detuning is decreased. Given that $\alpha_l L$ is about 0.93 at  $\delta/\Gamma = 7.5 $ for $OD=210$, corresponding to a single-pass transmission of only about $0.4$, the quasi-Kerr model seems to provide a useful guide to the true instability range even into regions where the absorption is far from negligible. The fit to the upper threshold curve is very good over the whole tuning range shown, because the absorption is strongly saturated in this region. The nonlinearity is saturated too, but the quasi-Kerr model fully accounts for that.

%

%
%

The similarities between the two-level quasi-Kerr and pure Kerr analyses can be exploited ``in reverse", to calculate thresholds for SFM pattern formation in Kerr media beyond the thin-medium models, for which some results (without detailed analysis) were reported in  \cite{labeyrie14}. Further, envelope curves can be calculated, so as to capture the range of thresholds afforded by varying the mirror displacement $D$, and to illustrate the thin-medium limit as discussed above.

\begin{figure}

 \includegraphics[scale=0.9,width=\columnwidth,trim=0 0 0 0,clip]{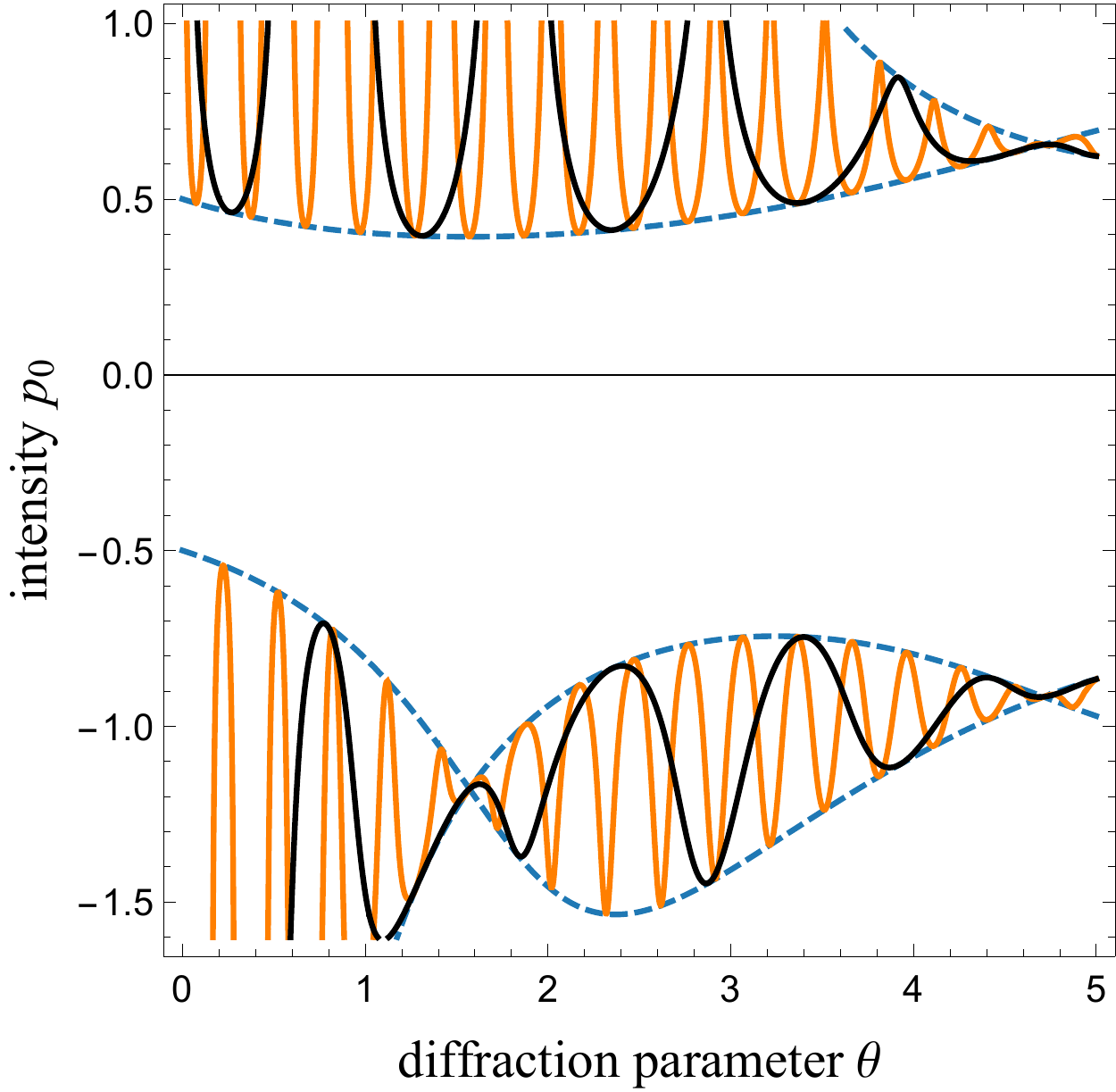}
\caption{(Color online) Threshold intensity (in units of $\alpha_l L \Delta p/2$) vs diffraction parameter $\theta =Q^2L/2k$. Blue dashed curves: Envelope curves calculated from (\ref{env}) for a Kerr medium with $h=0$. Positive and negative intensity values, respectively, correspond to self-focusing and self-defocusing Kerr media. Also threshold curves with a feedback mirror at negative effective distance from the end of the medium. Gray solid curves: $D=2.5$. Orange solid curves (with more wiggles): $D=10.0$. In both cases the threshold curves touch the envelope curves, and are confined by them.
}
 \label{fig:KG1env}
 \end{figure}

Figure~\ref{fig:KG1env} illustrates  this  for a Kerr medium with no grating term ($h=0$). Here two distances ($D=2.5,10$) are shown, and we begin to see how the faster oscillations of the threshold for larger mirror displacements allow a better exploration of the envelope, and thus potentially lower thresholds. For the self-focusing case, where the envelope has a minimum at finite $\theta$, we can see a transition of the lowest threshold  from the second-lowest-Q for $D=2.5$, to the sixth-lowest-Q band for $D=10$. Assuming that the dominant pattern is determined by the lowest threshold, we would expect that, as D is increased, the pattern period will slowly increase, and then suddenly drop back, in a sawtooth pattern. This phenomenon is indeed observed, as shown in Fig.~\ref{pol_Talbotscan}, where the dominance of the first Talbot ring for small $|D|$ is replaced by the second Talbot ring for larger $|D|$.

Conversely, for self-defocusing the lowest  threshold always decreases as $D$ is increased, so that the patterns with lowest threshold are found at large mirror displacements, and have large spatial scales, with pattern wavelength scaling like $\sqrt{d/k}$, as is well known from thin-medium theory \cite{firth90a}.
In contrast, CP thresholds for $h=0$ defocusing Kerr media decrease with increasing $Q$ (see e.g.\ \cite{Geddes1994}), so that phase-conjugate oscillation is the dominant instability.
This SFM advantage can be attributed to the ability of the feedback phase to compensate for both the diffractive and nonlinear phase shifts in the medium, which have the same sign for defocusing, and thus cannot cancel each other as they can for self-focusing. This no-grating Kerr case is also interesting in that the envelope curves cross, and hence the threshold curves must thread through the intersection (Fig.~\ref{fig:KG1env}). It follows that  the threshold is actually independent of mirror displacement at these crossings. Note that the threshold will normally be lower at a different diffraction parameter (as occurs in Fig.~\ref{fig:KG1env}), so observing the phenomenon would probably require isolating the specific wavenumber by Fourier filtering in the feedback loop \cite{pesch03}.

The finite limit for small diffraction, $\theta \to 0$, of the envelope is ($\pm 0.5$) in Fig. \ref{fig:KG1env}, and corresponds exactly to the thin-slice value \cite{firth90a}. It is clear from the above discussion that the small-$\theta$ region of the envelope can only be accessed for large $D$, and hence that the $\theta \to 0$ limit corresponds to $D \to \infty$, i.e. the thin-medium limit \cite{firth90a}. While previous thick-medium analyses \cite{Honda1996,kozyreff06} are valid in this limit, these authors did not explicitly consider it. The finite slope at $\theta=0$ means that the pattern-forming modes are not, in fact, threshold-degenerate when the medium thickness is taken into account. As is illustrated in Fig. \ref{fig:KG1env}, the multi-fractal patterns predicted in the thin-slice limit \cite{Huang2005} and dependent on mode-degeneracy are not expected to occur in practice, unless other mechanisms or devices are able to restore degeneracy. This effect of diffraction within the nonlinear medium was recognized earlier in \cite{kozyreff06}.



\section {Talbot Fans}

\begin{figure}
 \includegraphics[width=\columnwidth]{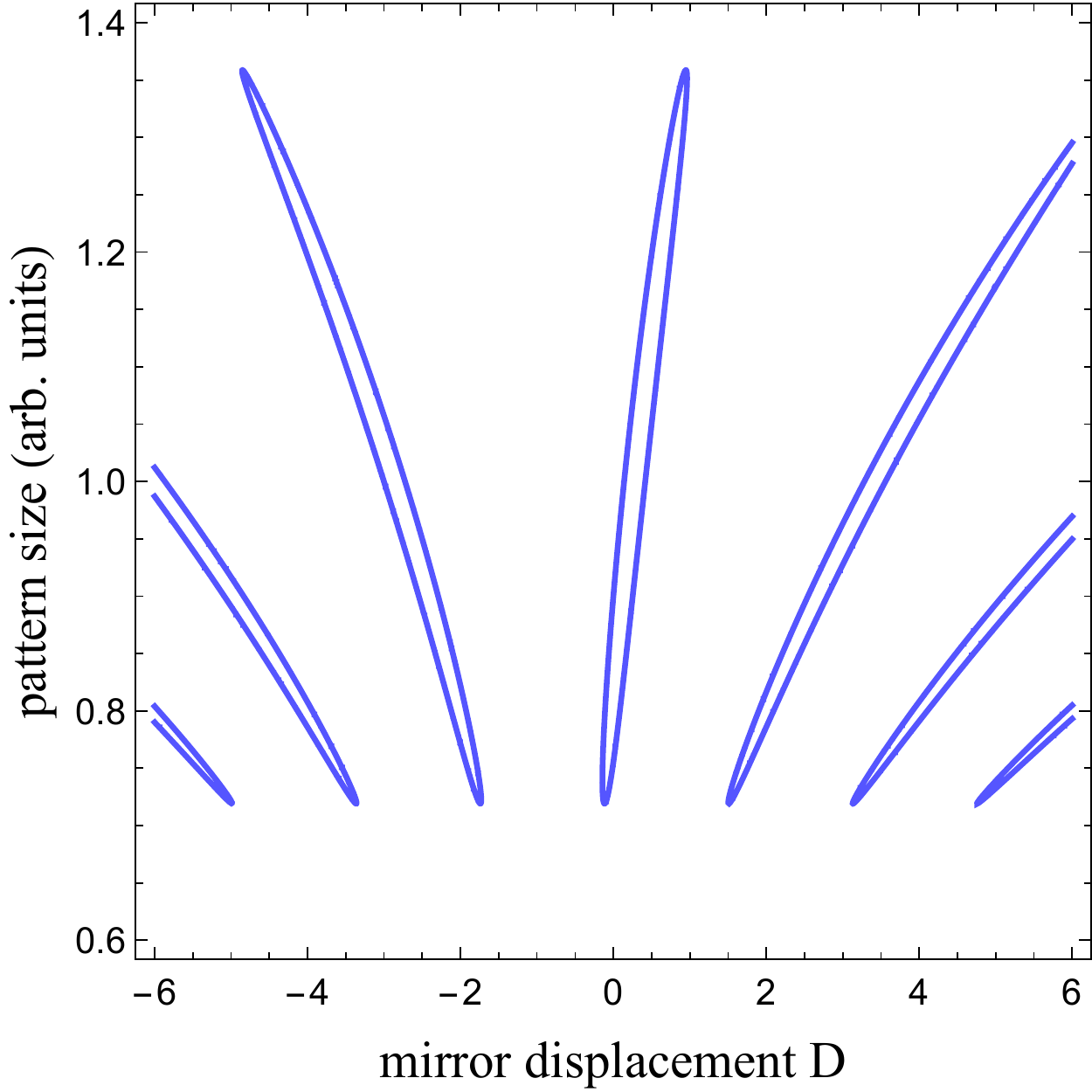}
\caption{ Pattern period (arb.\ units)  vs mirror displacement $D$ at fixed intensity $s=0.085$. Threshold curves  calculated from (\ref{2LSfbm}) for a two-level medium  described by $\hat{A}_{all}$, with $h=1$. The quasi-Kerr coefficient $\alpha_l L \Delta =13.94$, corresponding to blue detuning. For optical density 210 \cite{Camara2015}, this corresponds to detuning $\Delta = 2\delta/\Gamma = 15$.}
 \label{fig:2LSsize}
 \end{figure}

The above figures demonstrate how the threshold extrema move vs $\theta$ as mirror displacement $D$  is varied. An interesting and relevant way to examine this is to plot pattern scale ( $ \sim 1/\sqrt{\theta}$) vs $D$ for fixed intensity. This is demonstrated in Fig.~\ref{fig:2LSsize}, where the parameters are chosen to match those of  \cite{Camara2015}, and the intensity $s=0.085$ is just above the minimum threshold, so that the unstable regions appear as long narrow islands. The ``fan" shape of the island group is due to the Talbot effect: the threshold values satisfying (\ref{2LSfbm}) are evidently periodic in $\psi_D = D \theta$, which means that at fixed $\theta$ (size) and intensity, threshold values are periodic in $D$. This is particularly clear at the bottom of the fan in Fig. \ref{fig:2LSsize}, where the tips of the islands are equally-spaced in $D$. The Talbot periodicity is inversely proportional to $\theta$, which is why the islands fan out as the pattern scale increases (i.e. as $\theta$ decreases).

\begin{figure}
 \includegraphics[width=\columnwidth,trim=0 60 0 20,clip]{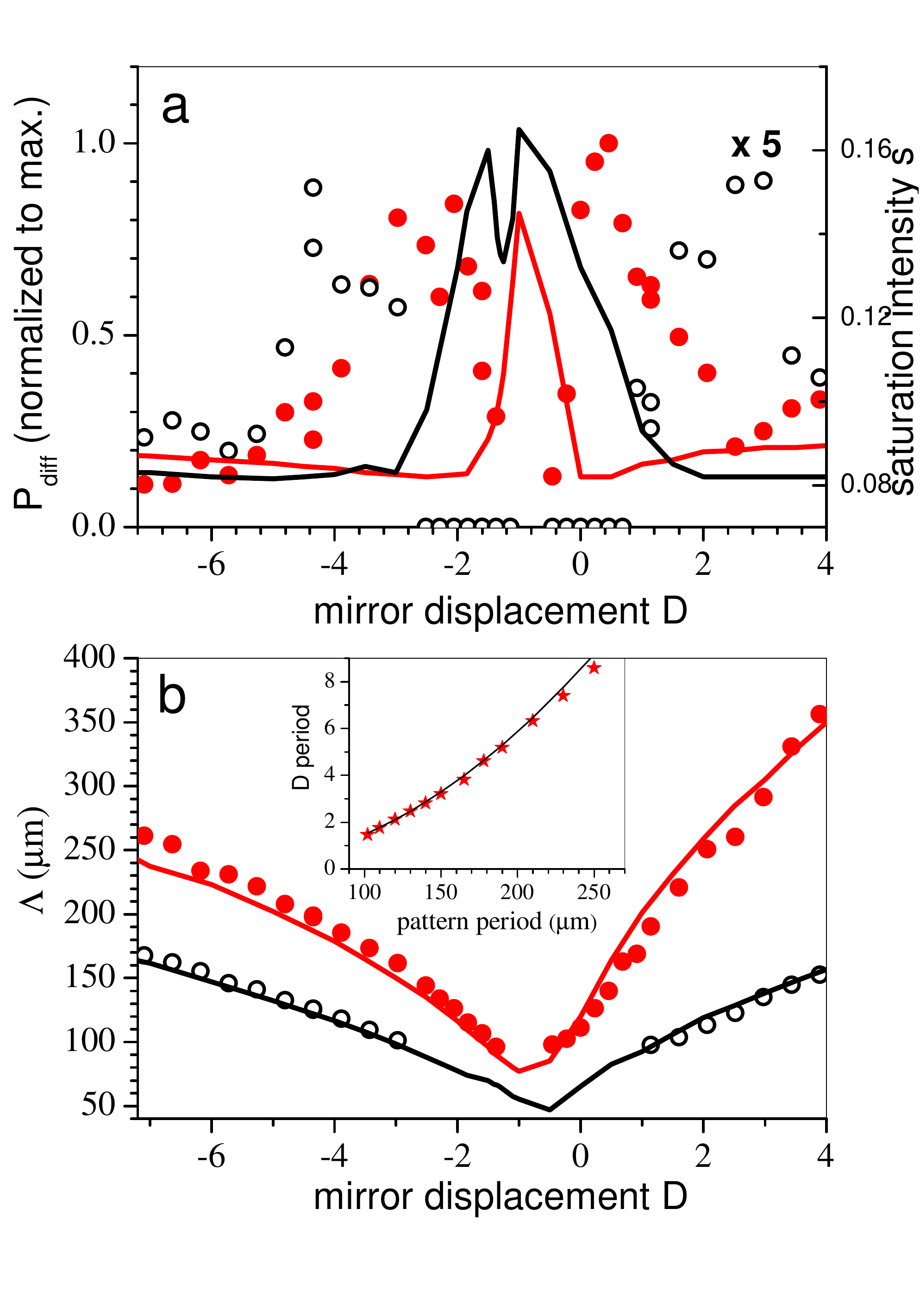}
\caption{(Color online) a) Diffracted power (experiment, left axis) and predicted threshold saturation intensity (theory, right axis) vs scaled mirror displacement $D$. The cloud thickness is $L=9$ mm. b) Pattern period  $\Lambda$ vs mirror displacement. In physical units, the $x$-axis corresponds to -60~mm to +40~mm measured from the center of the cloud. Parameters: blue detuning, $\Delta = 15$, see \cite{Camara2015}. The diffracted power is normalized to its maximal value. Red solid dots: experimental data for first Talbot balloon (lowest wavenumber), gray circles: experimental data for second Talbot balloon (next highest wavenumber excited, in a) enhanced by factor of 5). The red and gray curves are the corresponding theoretical predictions and are calculated from (\ref{2LSfbm}) using the all-grating two-level model.
 Inset: The measured $D$ period as a function of the pattern size (stars), together with the Talbot effect prediction (line). }
 \label{Talbotfan}
 \end{figure}

Such ``Talbot fans" are readily observed experimentally. The fan reported in \cite{Camara2015} is shown in Fig.~\ref{Talbotfan}, where the experimental data fit well to threshold data from  (\ref{2LSfbm}) using our two-level all-grating model based on $\hat{A}_{all}$. Fig.~\ref{Talbotfan} b) plots the pattern period against mirror displacement. Around $D\approx 0$ the lengthscale with the smallest wavenumber (largest period) is selected. At higher $|D|$, two lengthscales are found in the pattern. Both are in good agreement with the prediction from the theory. The inset shows excellent agreement between the measured and calculated $D$-periodicities. In the earlier optomechanical patterns paper \cite{labeyrie14}, there is a more limited fan, to which threshold data from (\ref{2LSfbm}) are fitted using a Kerr model ($h=0$, because the slow time scale allows atomic motion to wash out the longitudinal grating).

Fig.~\ref{Talbotfan} a) plots the power diffracted into the first and second unstable wavenumber obtained by integrating the measured far field intensity distributions over an annulus with the respective radius. We did not measure thresholds, but to a first approximation one can argue that the diffracted power increases with increasing distance to threshold and hence the measured data can be interpreted as indicators of inverted threshold curves. We compare them with the threshold curves obtained from the all grating quasi-Kerr model as the detuning is reasonably large and absorption not very important. As indicated in the discussion of  Fig.~\ref{Talbotfan} a), around $D\approx 0$, only the lowest wavenumber (i.e. the one from the first Talbot balloon) is excited. For a mirror within the medium ($D=-1 \dots 0$), the diffracted power is low and the predicted thresholds are high. For increasing $|D|$ threshold are predicted to fall dramatically and indeed well developed patterns, indicated by high diffracted power, are observed. For further increasing $|D|$ the theory predicts that the second Talbot balloon at higher wavenumber has the lowest threshold. Indeed excitation of this length scale is observed but it does not take over completely in the experimental data.

For a further investigation of the Talbot fan phenomenon we analyze a somewhat different experimental SFM situation in which optical pumping between Zeeman substates, rather than two-level electronic excitation, is the main nonlinearity \cite{grynberg94,scroggie96,leberre95b,aumann97}. Experimental parameters are an effective medium length of $L=3.2$ mm, beam intensity $I=18$ mW/cm$^2$ and detuning $\Delta = -14$. The homogenous solution is not saturated in this case \cite{ackemann01b}, so it is reasonable to compare the data to the length scales and threshold curves obtained from a self-focusing thick medium Kerr theory.

\begin{figure}
 \includegraphics[scale=0.9,width=\columnwidth,trim=0 10 0 0,clip]{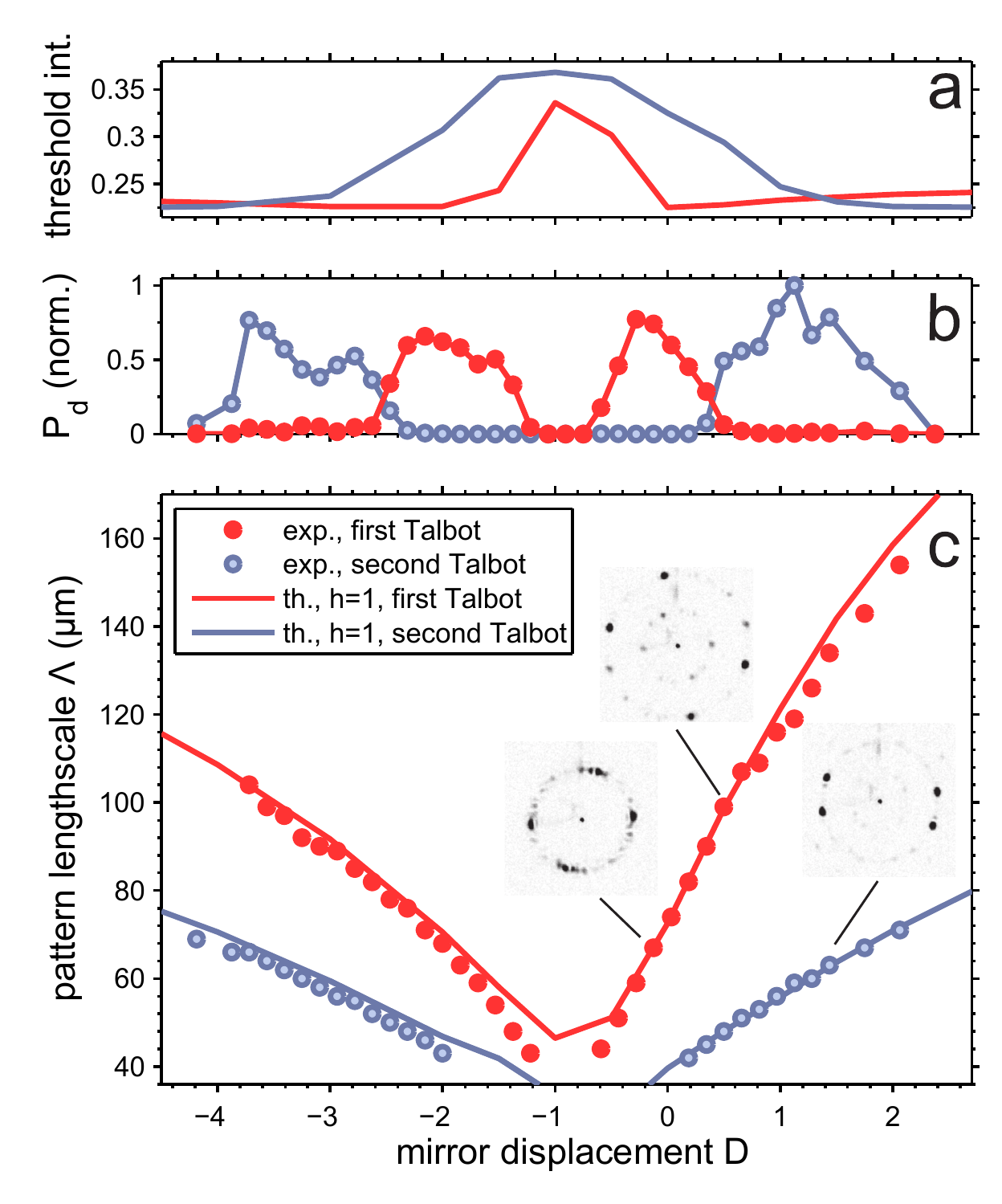}
\caption{ (Color online) a) Predicted threshold, b) experimentally observed diffracted power (normalized to its maximal value) and c) pattern period vs mirror displacement $D$. In unscaled parameters, the $x$-axis corresponds to -12.8~mm to +10.2~mm measured from cloud center. Parameters: effective medium length is $L=3.2$ mm, beam intensity $I=18$ mW/cm$^2$ and detuning $\Delta = -14$. Red solid dots: experimental data for first Talbot balloon (lowest wavenumber), blue circles: experimental data for second Talbot balloon (next highest wavenumber excited). The red and blue curves are the corresponding theoretical predictions and are calculated for a self-focusing Kerr medium with $h=1$ described by $\hat{A}_{Kerr}$. The insets show far field patterns obtained at the mirror positions indicated illustrating the length scale competition. }
\label{pol_Talbotscan}
\end{figure}

Experimental measurements of diffracted power and pattern lengthscale vs mirror displacement are shown in Fig.~\ref{pol_Talbotscan}. It is apparent that the behavior is very similar to the one observed for the electronic 2-level case in Fig.~\ref{Talbotfan}, but there is one crucial difference. For large enough $|D|$ ($D> 0.7$, $D<-2.5$) the power in the first Talbot ring is suppressed down to $3\times 10^{-3}$ relative to the second one, and the length scale of the second balloon takes over completely. This is in good, although not quantitative, agreement with the thick medium model as discussed earlier in connection with Figure~\ref{fig:KG1env}, though the transition is predicted to occur at somewhat larger $|D|$. Nevertheless, it is an important confirmation of the importance of the diffraction within the medium influencing length scale selection. In view of the fact that the atomic clouds have an approximately Gaussian density distribution and the theory assumes a rectangular distribution, quantitative deviations between theory and experiment are not surprising.

We note that a similar phenomenon was predicted in photorefractives \cite{schwab99,denz98}, in spite of different mechanism of non-linearity. However, the experimental observation of the essentially complete extinction of patterns with the smallest Talbot wavevector in favour of the second Talbot wavevector was not reported before in the literature, only the excitation of the second wavenumber (see Fig.~4 of \cite{Camara2015}, quantified in Fig.~\ref{Talbotfan} b) of this manuscript, for the two-level case and Fig.~7 of \cite{schwab99} for the photorefractive case). In hot atomic vapours, an early and not very systematic study \cite{ackemann94,ackemann96t} showed coexistence between the first Talbot wavevector and the second one for  $D\approx 2.4$ and between the  first Talbot wavevector and the third one for  $D\approx 3.9$. For even higher distances ($D\approx 8.3$) an excitation of a single, quite high order (five or six) Talbot wavevector was found. 
It should be noted that these results are influenced by atomic diffusion lifting the degeneracy present in the thin-slice model and a limited aspect ratio preventing patterns with the first Talbot wave vector for $D>4$.

Figures~\ref{Talbotfan} and \ref{pol_Talbotscan} indicate that a change of mirror displacement can drag the pattern period along qualitatively as in a diffractively thin medium but only up to a point. Then the system jumps back to a smaller length scale it seems to prefer, which can be changed again to some extent by changing mirror displacement. The origin of this behavior lies in the interaction between the threshold curves and the envelope as discussed before. For increasing $|D|$ the threshold curves move to lower $Q$ and have more wiggles in a certain range of $\theta$ on the envelope curve, which means they can explore more effectively the potentially lowest threshold condition. 

\begin{figure}
 \centering%
 \includegraphics[width=\columnwidth,clip]{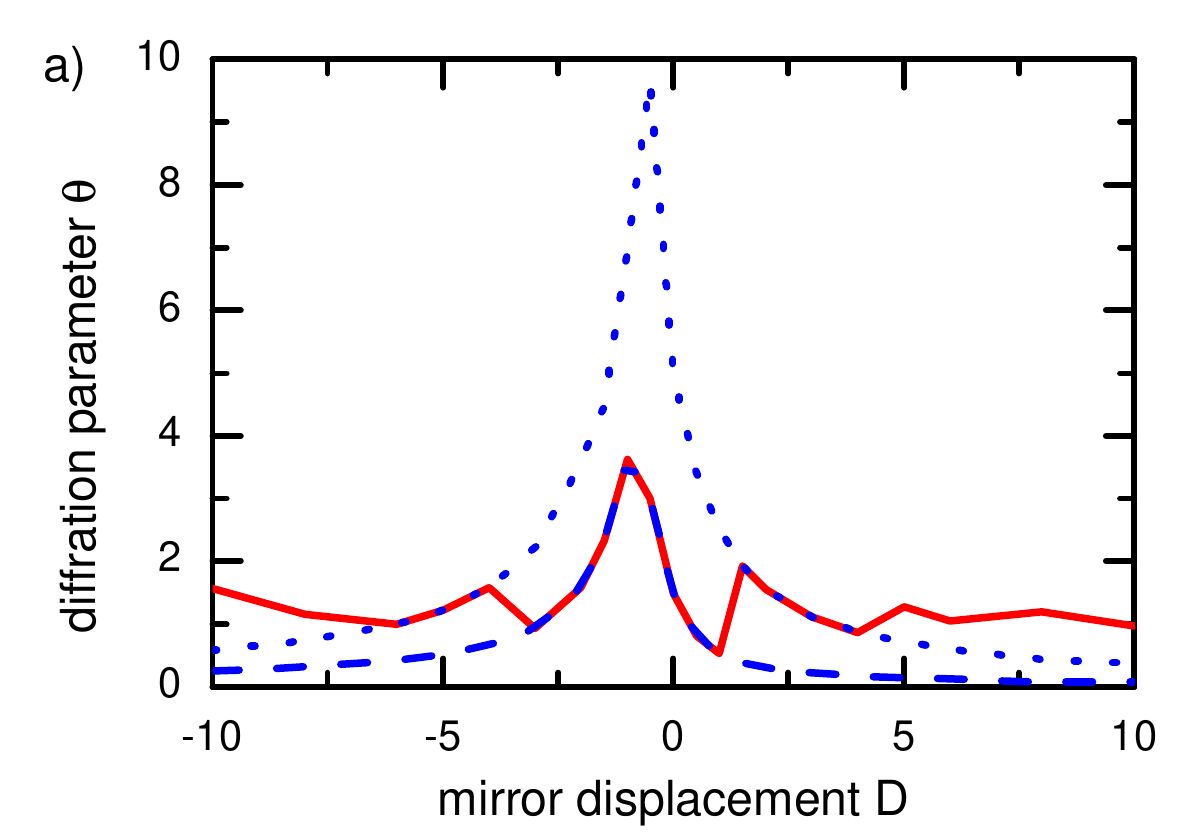}
\includegraphics[width=\columnwidth,clip]{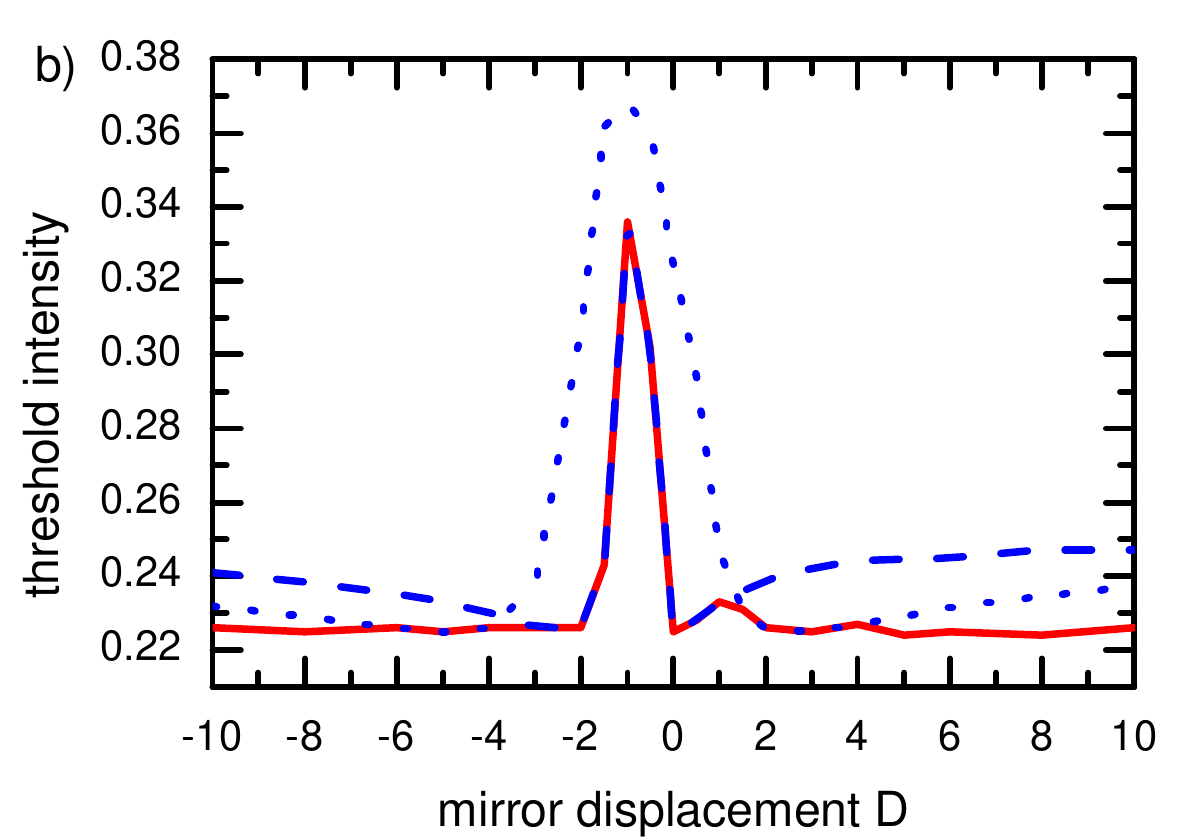}
\caption{(Color online) a) Pattern length scale (characterized by diffraction parameter $\theta$) and b) threshold intensity vs mirror displacement $D$ for a self-focusing Kerr medium with $h=1$ described by $\hat{A}_{Kerr}$. Red solid curve: minimum threshold, blue dashed curve: lowest wavenumber (first Talbot) balloon.}

\label{fig:D_thres}
 \end{figure}

Another way to illustrate this point is visualized in Fig.~\ref{fig:D_thres}. The red solid curve in Fig.~\ref{fig:D_thres} a) denotes the length scale of the minimum threshold mode vs mirror displacement. For $D=-3 \ldots 1$ it mirrors the first Talbot balloon, until it jumps to the second and follows it for $D=-6 \ldots -4$ and $D=1.5 \ldots 4$. Afterwards it jumps again and wiggles around a horizontal. The changes of lengthscale imply that the minimum of the envelope curve is at finite $\theta$ and the system is trying to stay close to this value as far as compatible with the specific boundary conditions, i.e. diffractive phase shift $\theta$ at the feedback distance $D$. 

This approach to the envelope curve is also nicely illustrated in the behavior of the threshold intensity vs $D$ (Fig.~\ref{fig:D_thres} b)), becoming nearly independent of distance for large mirror distances as the minimum of the envelope curve can be attained.

\section{Conclusion}
\label{conclusions}

In this paper we have undertaken a largely analytic investigation of thresholds and lengthscales for pattern formation in a saturable two-level medium, optically-excited close to resonance from one side, and with a feedback mirror to reflect and phase-shift the light fields after they have traversed the medium. In that scenario, we have established a number of results, in encouraging agreement with recent experimental results in several cases.

We have considered, and compared to experiment, the ``Talbot fan" characteristics which characterize the evolution of pattern scales as $D$ is varied, and explained observed sudden changes of scale in terms of mode competition in the neighborhood of the minimum possible (in $D$) threshold.

The additional degree of freedom offered by finite $D$ also implies an additional complexity in the analysis. We have shown, however, that thresholds are constrained by envelope curves to which the threshold curves are tangent, and along which they evolve as $D$ is varied. Hence important properties of the SFM system such as the minimum possible threshold, and the domains within which pattern formation is possible (or impossible) can be found, often analytically. Again, the envelope property is likely to be general, because it follows from the structure of the feedback boundary condition.

Importantly, the envelope functions enable a quantitative investigation of the limit $D \to\infty$, which correspond to diffraction in the medium being negligible compared to that in the feedback loop, i.e the thin-slice limit. We find that threshold values tend to precisely the thin-medium values, but with finite slope. As a consequence we have demonstrated that the degeneracy of the unstable modes predicted in thin-medium theory does not survive inclusion of finite medium length, even at lowest order.

Diffusive damping removing the degeneracy was introduced in the first treatments \cite{firth90a,dalessandro92} to  model carrier diffusion in semiconductors or elasto-viscous coupling in liquid crystals, which will make these media deviate from purely local Kerr media. In hot atom experiments \cite{grynberg94,Ackemann1994,ackemann95b} the thermal motion of the atoms, which can be modelled as diffusive motion under appropriate conditions \cite{Ackemann1994,ackemann95b}, will in tendency provide a stronger wash-out for transverse gratings at larger wavenumber and thus remove the degeneracy. In cold atoms this effect is not very strong and the finite medium thickness appears to be the main mechanism responsible for the emergence of a defined length scale in the investigations reported in Refs.~\cite{labeyrie14, Camara2015}. The possibility of a cut-off at high transverse wavenumbers due to the diffraction within the nonlinear medium (at least for some parameter combinations) was realized before in \cite{kozyreff06}.

In the specific context of the two-level nonlinearity we have analyzed different models to take account of wavelength scale (reflection) gratings in the steady-state susceptibility applicable to counterpropagation problems. We have found that models in which only the lowest-order ($2k$) gratings are considered predict a zero-order bistability as resonance is approached. This bistability disappears when all orders ($m\times 2k$) of gratings are included, and is therefore probably spurious. We have been able to develop models which include all grating orders, numerically for the fully-absorptive system and analytically in the quasi-Kerr and thin-medium limits, and have demonstrated reasonable agreement with experiment using these all-grating models.

In summary, we have developed a firm and systematic foundation for the analysis of the effects of in-medium diffraction, and of reflection gratings, in SFM  pattern formation. Though we have focused here on the saturable  two-level electronic nonlinearity, our approach and techniques have applicability across a wide class of nonlinearities. While our present analysis deals only with thresholds and steady-state instabilities, these are an important, and even essential, preliminary to more extensive numerical simulations, necessarily involving many additional parameters and many spatial and temporal scales. We already showed \cite{labeyrie14} that a simple thick-medium Kerr model gives useful insight into optomechanical SFM patterns, and in this work we have shown that a similar analysis helps understand important features of polarization-mediated SFM patterns in cold atoms.  Patterns in cold-atom clouds with laser irradiation and mirror feedback are proving to a be a very rich field, with diverse implications, and a secure basis for the interpretation of experimental results and the development of appropriate theoretical models is therefore very important.

\acknowledgments{The Strathclyde group is grateful for support by the Leverhulme Trust and an university studentship for IK by the University of Strathclyde. The Sophia Antipolis group is supported by CNRS, UNS, and R\'{e}gion PACA. The collaboration between the two groups was supported by Strathclyde Global Exchange Fund and CNRS. WJF also acknowledges sharing of unpublished work by M. Saffman. We are grateful to P. Gomes (first investigations of length scales like in Fig.~\ref{pol_Talbotscan} were done by him), A. Arnold and P. Griffin for experimental support, to G.R.M. Robb, G.-L. Oppo and R. Kaiser for fruitful discussions. }


\end{document}